\PassOptionsToPackage{numbers,sort&compress}{natbib}
\documentclass[preprintnumbers,amsmath,amssymb,prd, notitlepage,nofootinbib, superscriptaddress]{revtex4}

\usepackage{amsfonts,amssymb,amsmath}
\usepackage{gensymb}
\usepackage{color}
\usepackage{graphicx}
\usepackage{multirow}
\usepackage[utf8]{inputenc}
\usepackage{aas_macros}
\usepackage{hyperref}

\usepackage[normalem]{ulem}

\newcommand{\greaterthanapprox}{\mathrel{\vcenter{
  \offinterlineskip\halign{\hfil$##$\cr
    >\cr\noalign{\kern2pt}\sim\cr\noalign{\kern-2pt}}}}}
    
    \newcommand{\lessthanapprox}{\mathrel{\vcenter{
  \offinterlineskip\halign{\hfil$##$\cr
    <\cr\noalign{\kern2pt}\sim\cr\noalign{\kern-2pt}}}}}

    \newcommand{\Npix}{N_{\mathrm{pix}}}

\newcommand{\lb}{\left(}
\newcommand{\rb}{\right)}

\newcommand{\be}{\begin{equation}}        
\newcommand{\ee}{\end{equation}}

\begin{document}

\title{Signal-preserving CMB component separation with machine learning}

\author{Fiona McCarthy}
\email{fmm43@cam.ac.uk}
\affiliation{DAMTP, Centre for Mathematical Sciences, Wilberforce Road, Cambridge CB3 0WA, UK}
\affiliation{Kavli Institute for Cosmology Cambridge, Madingley Road, Cambridge, CB3 0HA, UK}
\affiliation{Center for Computational Astrophysics, Flatiron Institute, New York, NY, USA 10010}

\author{J.~Colin Hill}
\affiliation{Department of Physics, Columbia University, New York, NY, USA 10027}

\author{William R.~Coulton}
\affiliation{Kavli Institute for Cosmology Cambridge, Madingley Road, Cambridge, CB3 0HA, UK}
\affiliation{DAMTP, Centre for Mathematical Sciences, Wilberforce Road, Cambridge CB3 0WA, UK}
\affiliation{Center for Computational Astrophysics, Flatiron Institute, New York, NY, USA 10010}

\author{David W.~Hogg}
\affiliation{Center for Computational Astrophysics, Flatiron Institute, New York, NY, USA 10010}
\affiliation{Max-Planck-Institut fur Astronomie, Konigstuhl 17, D-69117 Heidelberg, Germany}
\affiliation{Center for Cosmology and Particle Physics, Department of Physics, New York University, 726 Broadway, New York, NY 10003, USA
}
\date{\today}

\begin{abstract}

Analysis of microwave sky signals, such as the cosmic microwave background, often requires component separation using multi-frequency methods, whereby different signals are isolated according to their different frequency behaviors. Many so-called ``blind'' methods, such as the internal linear combination (ILC), make minimal assumptions about the spatial distribution of the signal or contaminants, and only assume knowledge of the frequency dependence of the signal.  The ILC produces a minimum-variance linear combination of the measured frequency maps. In the case of Gaussian, statistically isotropic fields, this is the optimal linear combination, as the variance is the only statistic of interest. However, in many cases the signal we wish to isolate, or the foregrounds we wish to remove, are non-Gaussian and/or statistically anisotropic (in particular for the case of Galactic foregrounds). In such cases, it is possible that machine learning (ML) techniques can be used to exploit the non-Gaussian features of the foregrounds and thereby improve component separation. However, many ML techniques require the use of complex, difficult-to-interpret operations on the data. We propose a hybrid method whereby we train an ML model using only combinations of the data that \textit{do not contain the signal}, and combine the resulting ML-predicted foreground estimate with the ILC solution to reduce the error from the ILC. We demonstrate our methods on simulations of extragalactic temperature and Galactic polarization foregrounds, and show that our ML model can exploit non-Gaussian features, such as point sources and spatially-varying spectral indices, to produce lower-variance maps than ILC --- e.g., reducing the variance of the B-mode residual by factors of up to 5 --- while preserving the signal of interest in an unbiased manner.  Moreover, we often find improved performance even when applying our ML technique to foreground models on which it was not trained.

 \end{abstract}
\maketitle

\section{Introduction}

When we observe the sky, we detect emission from many different sources, including extragalactic and Galactic. A notable example is the measurement of the cosmic microwave background (CMB) radiation at millimeter wavelengths, at which we detect not only the primary CMB but also the Sunyaev--Zel'dovich effect (sourced by the scattering of the CMB from free electrons in late-Universe structures); the cosmic infrared background (CIB); radio emission from extragalactic sources; and various types of radiation from our own Galaxy, including thermal dust radiation, synchrotron, and free-free emission. Much information can be extracted from these signals, but it is necessary to be able to separate them reliably. The science goals of current and upcoming CMB anisotropy experiments --- such as the search for evidence of primordial gravitational waves by the BICEP Array~\cite{2018SPIE10708E..07H}, Simons Observatory (SO)~\cite{2019JCAP...02..056A}, CMB-S4~\cite{2016arXiv161002743A}, and LiteBIRD~\cite{2023PTEP.2023d2F01L} --- will depend critically on our ability to separate the CMB from foreground signals.

To separate them, it is common to use the fact that the different sources exhibit different frequency dependence, quantified through their spectral energy distributions (SEDs). One can then observe the sky at multiple wavelengths and use this frequency information to isolate one (or multiple) component(s) of interest. Various methods exist for doing so, including methods like \texttt{Commander}~\cite{2008ApJ...676...10E}, which constrain models for various components in a Bayesian manner by sampling foreground templates and spectral behavior in a pixel-by-pixel  manner; and ``blind'' methods like \texttt{NILC} (Needlet ILC)~\cite{2009A&A...493..835D};  \texttt{MILCA} (modified ILC algorithm); \texttt{SEVEM} (spectral estimation via expectation maximization)~\cite{2003MNRAS.345.1101M}; and \texttt{SMICA} (spectral matching independent component analysis)~\cite{2008ISTSP...2..735C}, which all use minimal modeling assumptions to form (quasi)-linear combinations of the frequency maps to produce a map of a component of interest whose frequency behavior is well-understood.  Several of these algorithms are variations of the internal linear combination (ILC) method.

In the case of ILC methods, the coefficients of the linear combinations are chosen to \textit{minimize the variance} of the final map, computed on some domain. Such a choice has various advantages: 
\begin{enumerate}
\item it is  analytically tractable given an  estimate of the  covariance matrix of the data (which is estimated directly from the data in ILC --- hence the nomenclature ``internal''); 
\item it is possible to \textit{exactly} retain the signal (if its frequency dependence is known a priori, and if the foregrounds are uncorrelated with the signal); and 
\item it is the \textit{optimal} choice of linear combination of the maps in the case where the data are Gaussian and isotropic. 
\end{enumerate}
These advantages and tractability have led to such methods being widely used on CMB data in various forms~\cite{1992ApJ...396L...7B,2003ApJS..148...97B,2007ApJS..170..335P,2009A&A...493..835D,2014JCAP...02..030H,2016A&A...594A..22P,2020PhRvD.102b3534M,2022ApJS..258...36B,2022MNRAS.509..300T,2023arXiv230510193C,2023arXiv230701043M,2023arXiv230701258C}, including on \textit{COBE}~\cite{1992ApJ...397..420B}, \textit{WMAP}~\cite{2003ApJ...583....1B}, \textit{Planck}~\cite{2020A&A...641A...1P}, and Atacama Cosmology Telescope (ACT)~\cite{2016JLTP..184..772H}  (and various combinations thereof), to construct estimates both of the CMB+kinetic Sunyaev--Zel'dovich (kSZ) signal and of the thermal Sunyaev--Zel'dovich (tSZ) signal.\footnote{These two components have frequency dependences that are effectively exactly known,  with the CMB and kSZ described by a blackbody at temperature $T=2.725 \,\, \mathrm{K}$~\cite{1996ApJ...473..576F} and the (non-relativistic) tSZ distortion well-understood analytically~\cite{1970Ap&SS...7....3S}.} However, it is interesting to develop improved algorithms beyond the Gaussian, isotropic case.  For example, the \texttt{NILC} algorithm was designed to allow flexibility for anisotropic foregrounds by minimizing the variance separately on different spatially- and harmonically-localized domains; however, this remains optimal only when the data are isotropic and Gaussian within these domains.

Non-Gaussian fields are of course relevant for CMB experiments. In particular, the Galactic foregrounds that are already known to dominate over the (potential) primordial B-mode polarization signal sourced by inflationary gravitational waves~\cite{1997PhRvL..78.2058K,1997PhRvL..78.2054S} are highly non-Gaussian and anisotropic, with emission stronger towards the Galactic center, as well as displaying spatially varying SEDs. These foregrounds represent one of the largest sources of difficulty in constraining the tensor-to-scalar ratio  $r$~\cite{2014JCAP...08..039F,2015PhRvL.114j1301B}. Additionally, in the \textit{small-scale}, \textit{high-resolution} frontier of CMB temperature secondary anisotropies, the \textit{extragalactic} foregrounds that dominate over the CMB are non-Gaussian, with the CIB being composed of {point sources that cluster on large scales but are Poisson-like on small scales,} and the tSZ sources also being Poisson-distributed (on small scales). {The kSZ anisotropies are also non-Gaussian.} Separating these signals will be necessary to fully exploit the information in the ground-based experiments (such as ACT~\cite{2016JLTP..184..772H}, SPT~\cite{2011PASP..123..568C,2014SPIE.9153E..1PB}, SO~\cite{2019JCAP...02..056A}, and CMB-S4~\cite{2016arXiv161002743A}, as well as the more futuristic CMB-HD~\cite{2019BAAS...51g...6S}) that will be making measurements in this regime.

The ILC estimate can be easily computed. {However, in the presence of higher-order statistics that we may wish to minimize instead of the variance,} we lose the analytic tractability that is so attractive in the simpler case. Such a regime is perhaps an obvious candidate to exploit the recent explosion in machine learning (ML) tools in many areas of science. Indeed, several proposals for component separation using various flavors of machine learning {or higher-order statistical techniques} have been proposed, including for the CMB case~\cite{2020ApJ...903..104P,2022A&A...666A..89C,2023arXiv231007590C,2022ApJS..260...13W,2023arXiv230212378A,2022arXiv221202847B,2021A&A...649L..18R,2024A&A...681A...1A,2023arXiv231016285H,2014PhRvL.113s1303K,2018MNRAS.479.5577P} and other astrophysical applications such as foregrounds for 21 cm cosmology surveys~\cite{2021JCAP...04..081M,2023arXiv231006518S}. Generally, such ML tools require training on simulations {(with the exception of methods that take a self-supervised approach such as that in Ref.~\cite{2022arXiv221202847B})}. Therefore, they necessarily depend on the accuracy of the simulations used in training. Given the complicated nature of the physics sourcing the signals we are concerned with, the ``black box'' of an ML algorithm is something that one must be wary of. {In particular, it is challenging to simulate the Galactic dust and synchrotron emission relevant for polarization component separation.} In this work, we present an ML component-separation method that inherits many of the desirable qualities of the ILC case: it is explicitly constructed to be \textit{linear} in the signal of interest, and \textit{unbiased} to it. We impose these constraints by constructing linear combinations of the data that are necessarily \textit{insensitive} to the signal of interest (e.g., by taking differences of maps from neighboring frequency channels) and only passing these through our ML model, to build non-linear estimates of the \textit{foregrounds}, which we then subtract from an estimate of the signal. We demonstrate our method by training on two cases: non-Gaussian simulations of high-resolution temperature observations, and non-Gaussian and anisotropic large-scale B-mode polarization observations.

The paper is organized as follows. We present and discuss the ILC framework and solution in Section~\ref{sec:ILC}. In Section~\ref{sec:signal_preserving_vs_direct} we present our nonlinear, signal-preserving approach. We demonstrate our methods on both noiseless and  SO-like extragalactic temperature measurements of the CMB and tSZ fields in Section~\ref{sec:small_scale_results}. We demonstrate our approach on Galactic foregrounds and the CMB B-mode signal in Section~\ref{sec:large_scale_bmode_results}. We conclude in Section~\ref{sec:conclusion}. Note that, throughout, for all of our ML models, we use a UNET~\cite{2015arXiv150504597R}-like architecture{, as described in Appendix~\ref{app:architecture}}. 

\section{Internal Linear Combination: ILC}\label{sec:ILC}

In Section~\ref{sec:intro_ILC} we introduce the ILC framework. In Section~\ref{sec:unbiased_ILC}  we discuss its performance on a signal of interest:  a simulated extragalactic CMB temperature observation with SO-like frequency coverage and resolution. We explicitly demonstrate the preservation of the signal in an ILC.

\subsection{Introduction to the ILC framework}\label{sec:intro_ILC}

Consider multiwavelength observations of the same patch of sky in $N^{\mathrm{chann}}$  frequency channels. The temperature in pixel $p$ in channel $i$ is given by
\be
T_i(p) = T_i{}^{\mathrm{coi}}(p) + F_i(p) + N_i(p),
\ee
where $ T_i{}^{\mathrm{coi}}(p) $ is the temperature of a ``component of interest'' in pixel $p$; $F_i(p)$ is the signal contribution from foregrounds/backgrounds (extragalactic and Galactic); and $N_i(p)$ is the noise in pixel $p$, channel $i$. If the component of interest is the CMB, and one is working in linearized differential thermodynamic temperature units (i.e., CMB temperature units), then $T_i{}^{\mathrm{coi}}(p)=T^{CMB}(p)$ in all channels (note the lack of frequency dependence, which is by construction). Other possible components of interest, such as the tSZ signal, depend on $i$; while one can always divide this out and work in the relevant units for the signal of interest, it can also be helpful to define an SED vector $a_i$ such that
\be
T_i{}^{\mathrm{coi}}(p)  = a_i T^{\mathrm{coi}}(p) \,.
\ee
Note that this assumes \textit{no frequency decorrelation} of the signal of interest, which is a valid assumption for the CMB and tSZ signals, but not necessarily for others. For the CMB in CMB temperature units, $a_i$ is a vector of ones. 

The idea behind the ILC is to construct a linear combination of the frequency maps so as to build an estimate of the signal of interest:
\be
\hat T^{\mathrm{coi}}(p) = \sum _i w_i T_i(p)
\ee
where $w_i$ are weight coefficients (which may or may not depend on $p$) and $\hat T^{\mathrm{coi}}(p)$ is the resulting estimate of the signal of interest. In order for this to be a true estimate of the signal of interest, we require that
\be
\sum _i a_i w_i = 1 \,.
\ee
This condition ensures that the linear combination is unbiased with respect to the component of interest. 

We further want to recover, in some sense, the ``best'' unbiased estimate of the component of interest; one could define this in many ways, but the standard is to require that the linear combination is that which \textit{minimizes the variance} of the resulting map. Given knowledge of the covariance matrix of the data $C$ (in frequency space), we can use Lagrange multipliers to derive the appropriate coefficients~\cite{2004ApJ...612..633E}:
\begin{equation}
w_i = \left[\left( a^T  C^{-1} a\right)^{-1}\right]\left[a^T\left( C^{-1}\right)\right]_{i}.
\end{equation}

In the ILC approach, we use the data itself to gain knowledge of $C$, by directly measuring the covariance matrix $\hat C$, over some domain $\mathcal D$ of the data. However, the use of the data to calculate the weights leads to the well-known ``ILC bias'' where chance correlations in the data bias the estimate of the signal~\cite{2009A&A...493..835D}.  To mitigate this bias, it is necessary to use a large enough domain to calculate the ILC weights {(also see the ILC-bias-reduction techniques of~\cite{2023arXiv230701258C}, where the covariance matrix is computed on a domain that excludes the region it will be applied to)}. Alternatively, a theoretical covariance matrix (or one calculated from simulations) can be used, to incur no ILC bias, as was done in~\cite{2022ApJS..258...36B}, although the linear combination is no longer ``internal.''

There are several domains one can choose on which to estimate the covariance matrix and hence perform the ILC.  Common choices are  spatial (real-space) domains, harmonic domains (HILC), and needlet domains (NILC)~\cite{2009A&A...493..835D}, the last of which combines some of the aspects of the first two. ILC has been applied to CMB datasets for many years; see~\cite{1992ApJ...396L...7B} for an application to \textit{COBE} data and~\cite{2003ApJS..148...97B,2004ApJ...612..633E} for \textit{WMAP} data
 (all spatial ILC), as well as later applications of NILC on \textit{WMAP} data~\cite{2009A&A...493..835D}, \textit{Planck} data~\cite{2016A&A...594A..10P,2020A&A...641A...4P} (including to isolate the tSZ effect~\cite{2016A&A...594A..22P,2022MNRAS.509..300T,2023MNRAS.526.5682C,2024PhRvD.109b3528M} and other signals like the spatially varying $\mu$-distortion~\cite{2022MNRAS.515.5847R}), and ACT data~\cite{2023arXiv230701258C}, and HILC on \textit{Planck} data~\cite{2014JCAP...02..030H} and ACT data~\cite{2020PhRvD.102b3534M}.
 We will briefly discuss spatial and harmonic ILC (HILC) below.

\subsubsection*{Spatial ILC}

One of the simplest domains on which to perform the ILC is the spatial domain of the data. In this case, the covariance matrix $\hat C$ is just given by
\be
\hat C_{ij} = \left<T_i(p)T_j(p)  \right>=\sum _p T_i(p)T_j(p) \,,
\ee
assuming that the data have zero mean.  Note that, in this case, $ C_{ij}$ does not necessarily have pixel dependence: we can measure the same covariance (and thus apply the same ILC weights) over the entire domain of the map. However, we can allow for some spatial dependence by defining smaller pixel-dependent real-space domains $\mathcal D_p$ and measuring the covariance within each smaller domain $\mathcal D_p$.

In most cases in this work, we will consider spatial ILC; in particular, we take the spatial ILC as our starting point for our frequency cleaning methods. We will compare our final results to the harmonic ILC (HILC), which is performed in the domain of the spherical harmonics of the data, which is the appropriate basis when the components have different scale dependence and are spatially isotropic. We introduce HILC below.

\begin{figure}[t!]
\includegraphics[width=0.24\textwidth]{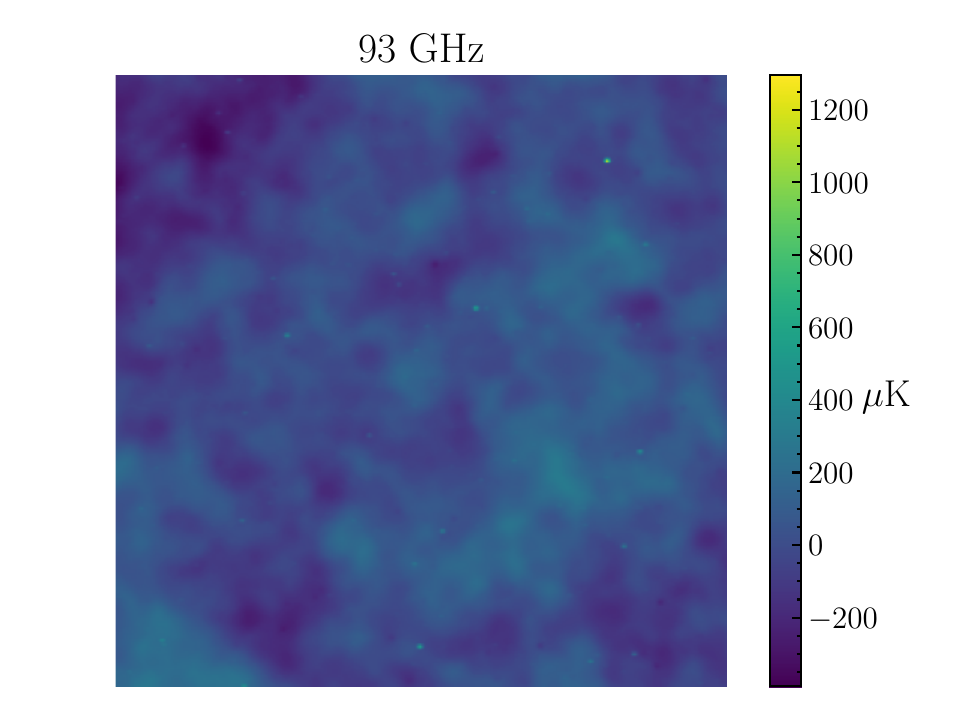}
\includegraphics[width=0.24\textwidth]{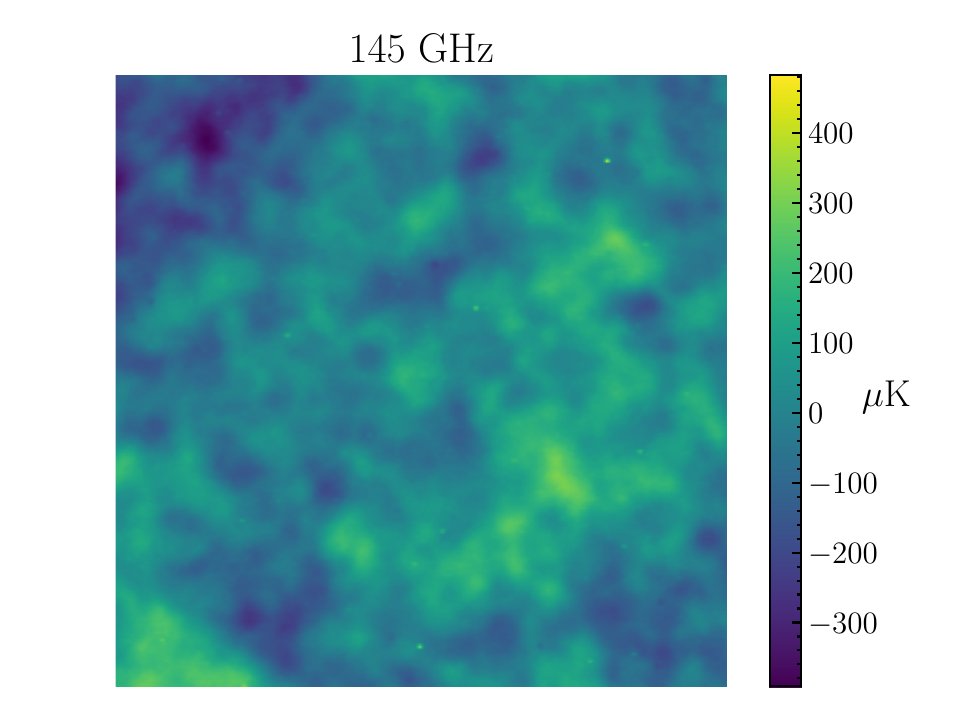}
\includegraphics[width=0.24\textwidth]{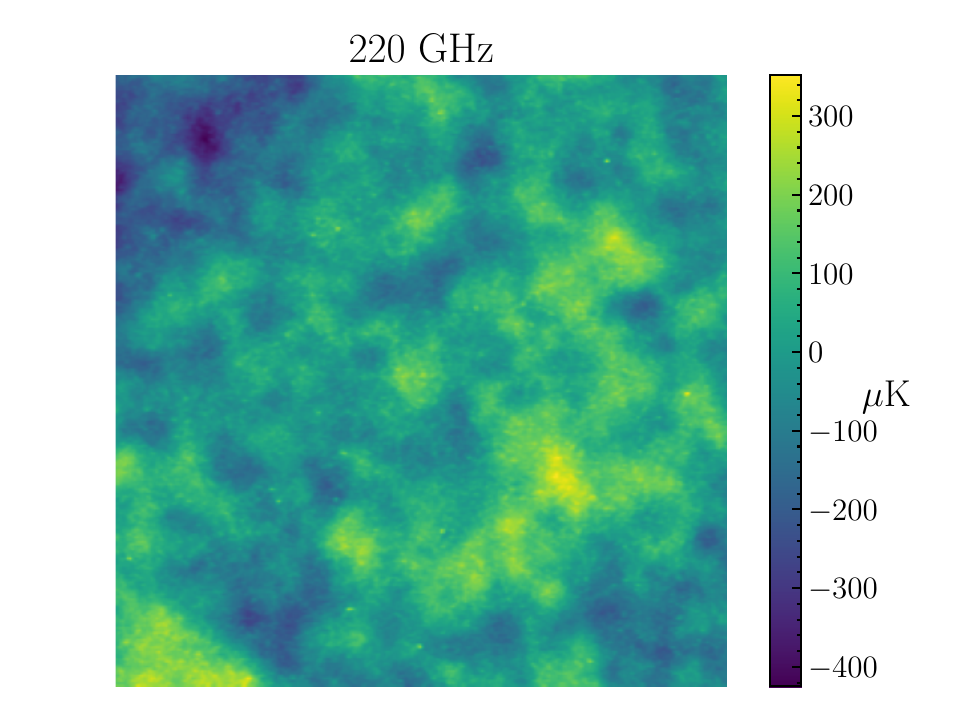}
\includegraphics[width=0.24\textwidth]{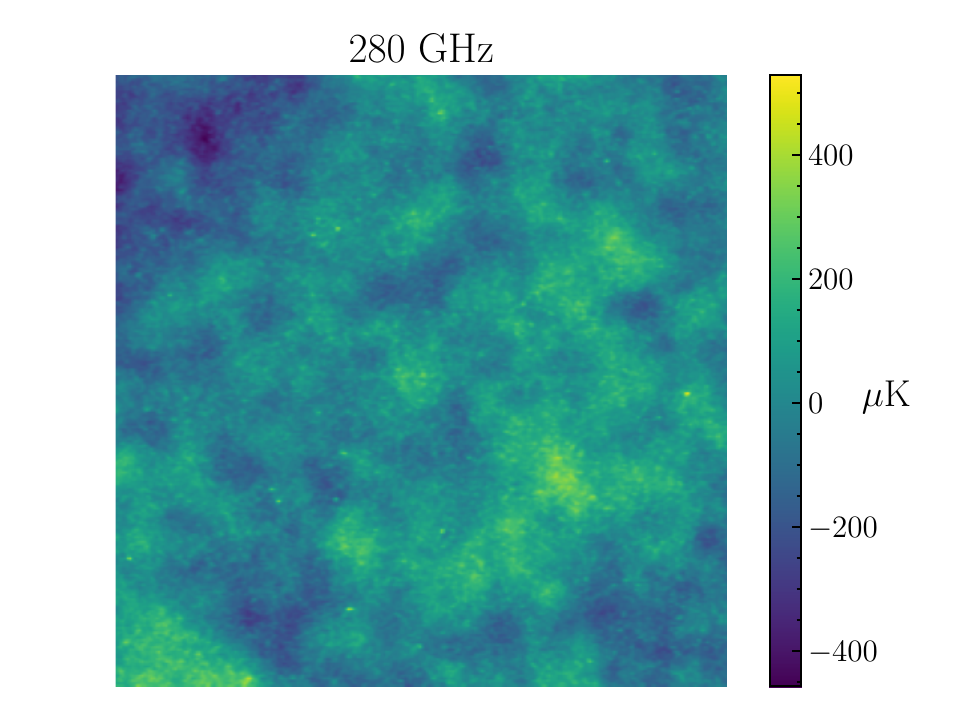}\\

\includegraphics[width=0.24\textwidth]{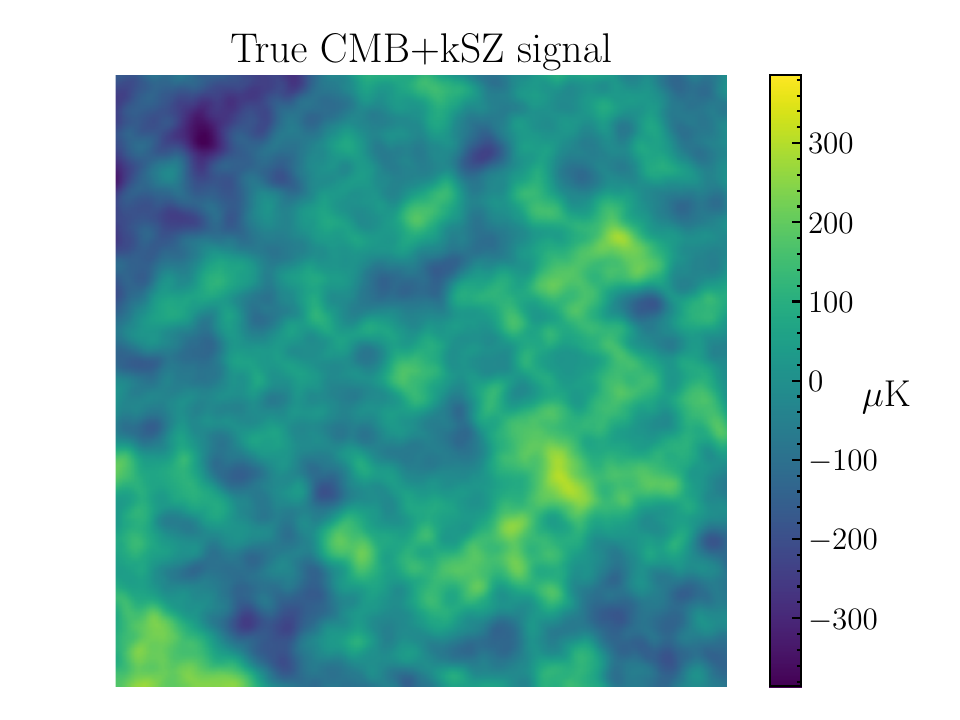}
\includegraphics[width=0.24\textwidth]{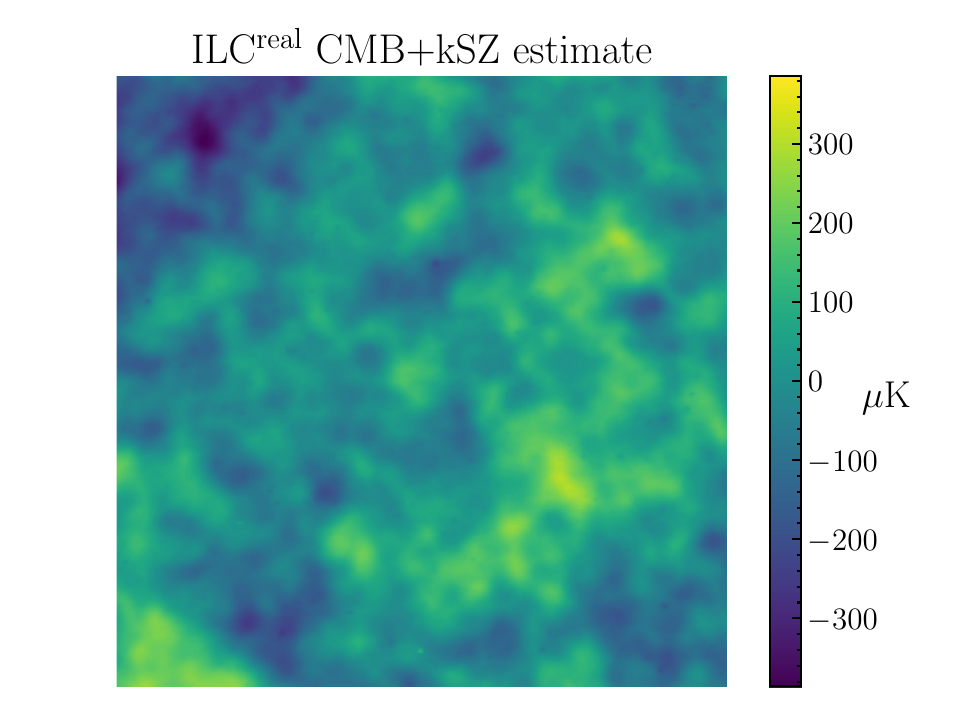}
\includegraphics[width=0.24\textwidth]{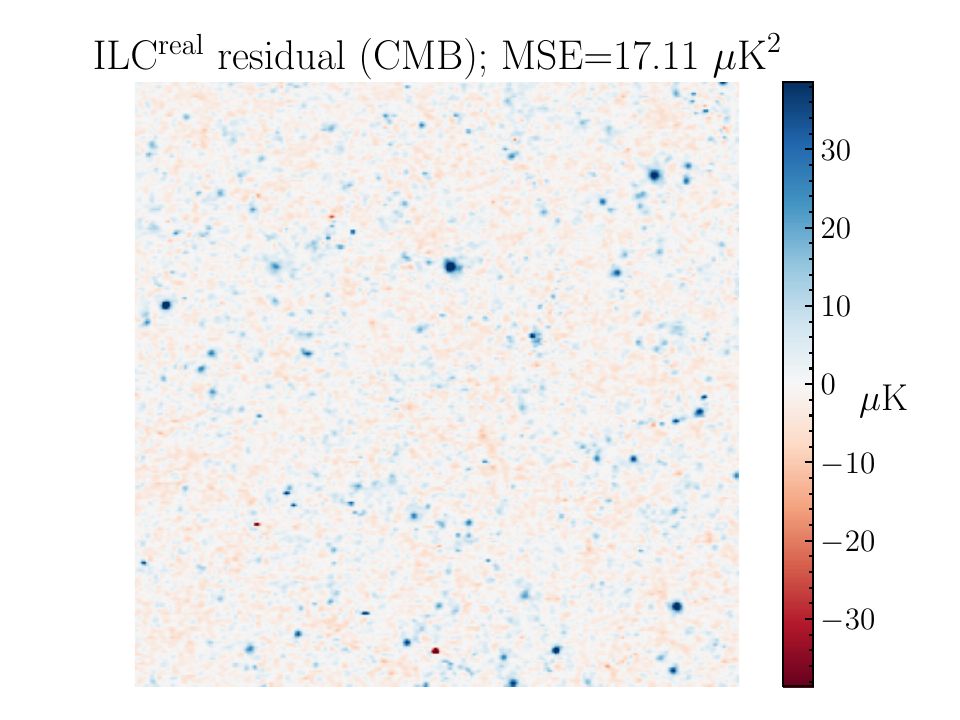}
\includegraphics[width=0.24\textwidth]{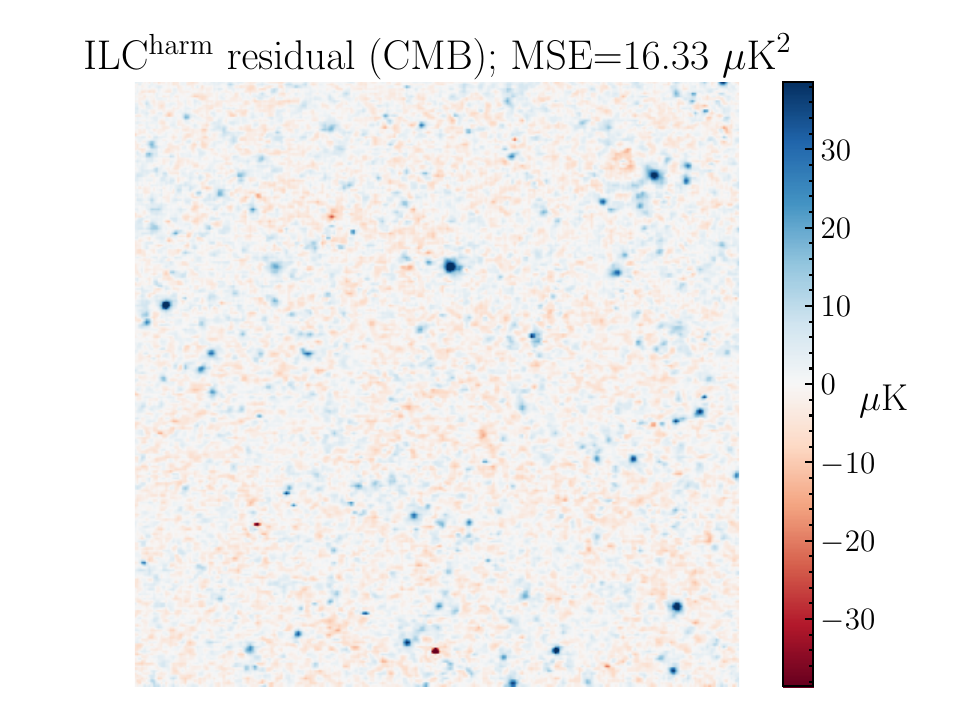}\\

\includegraphics[width=0.24\textwidth]{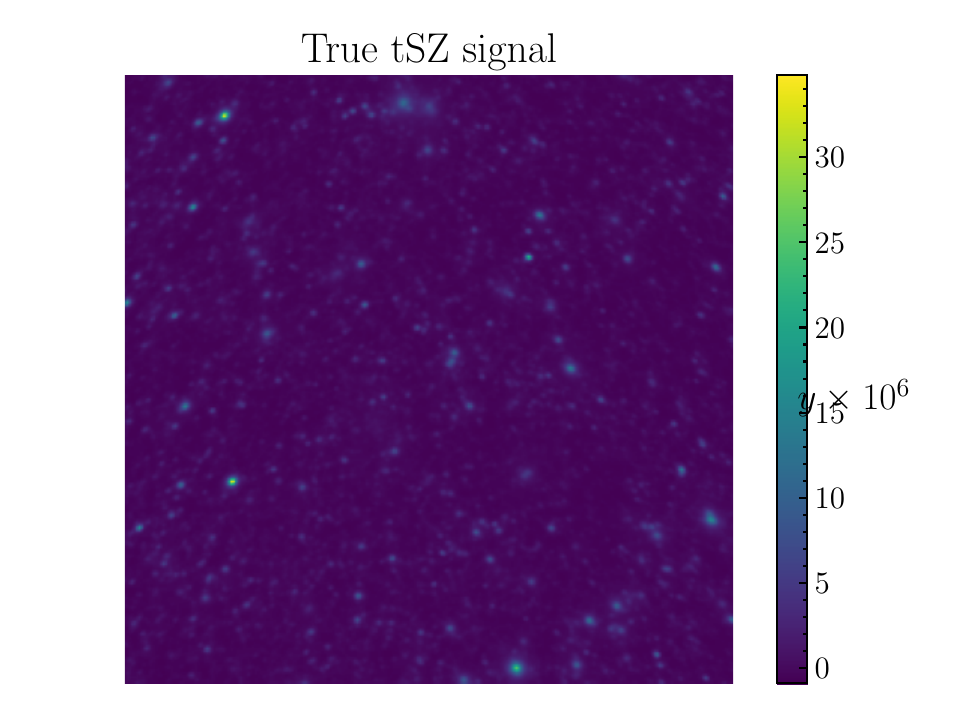}
\includegraphics[width=0.24\textwidth]{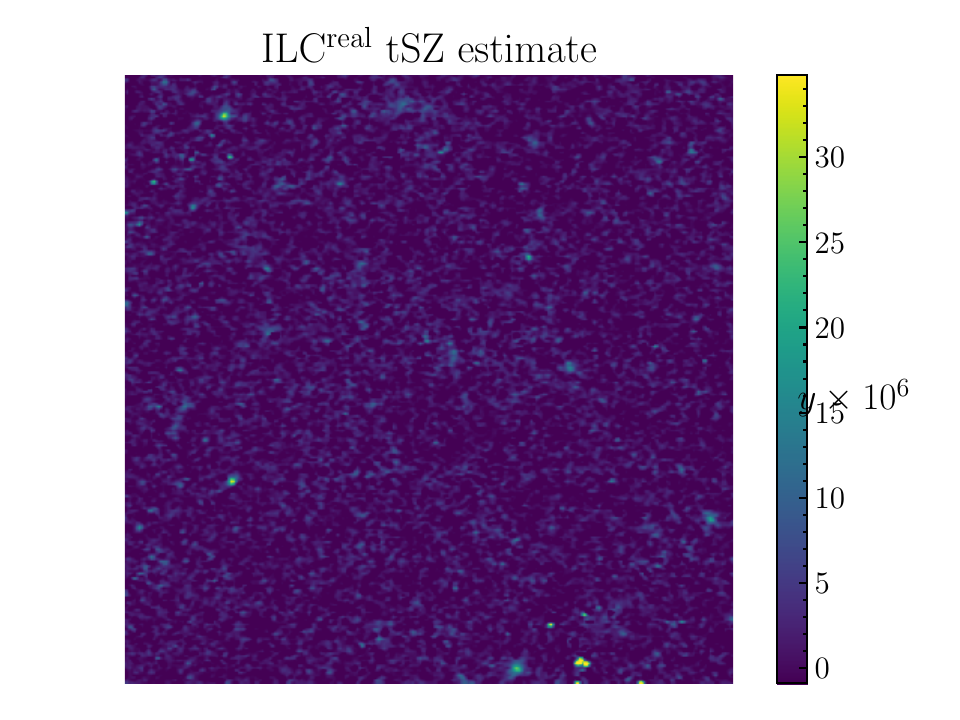}
\includegraphics[width=0.24\textwidth]{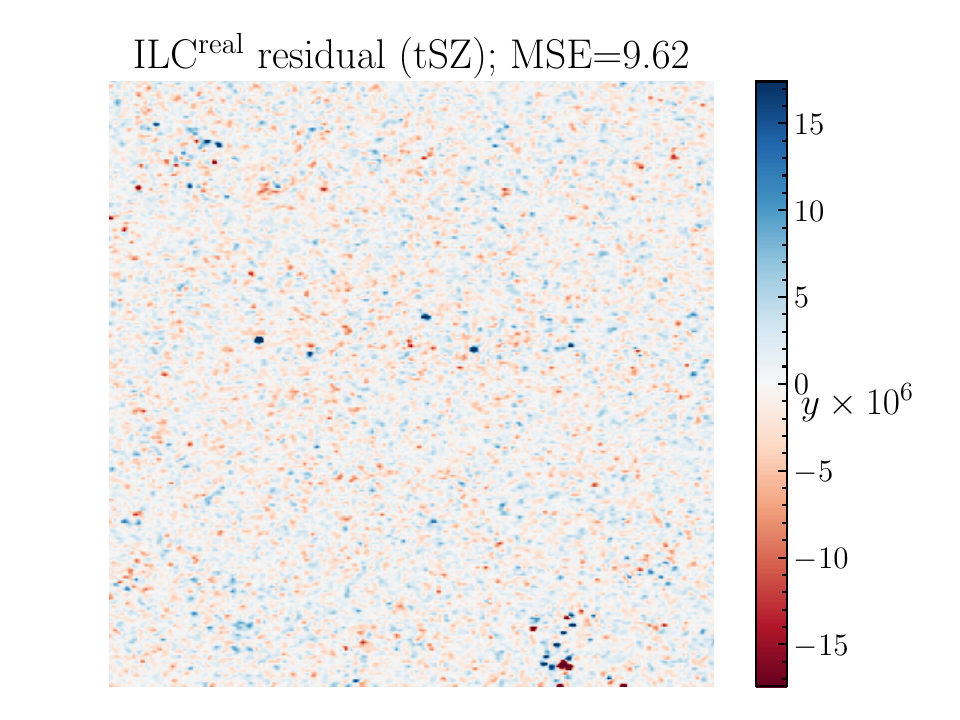}
\includegraphics[width=0.24\textwidth]{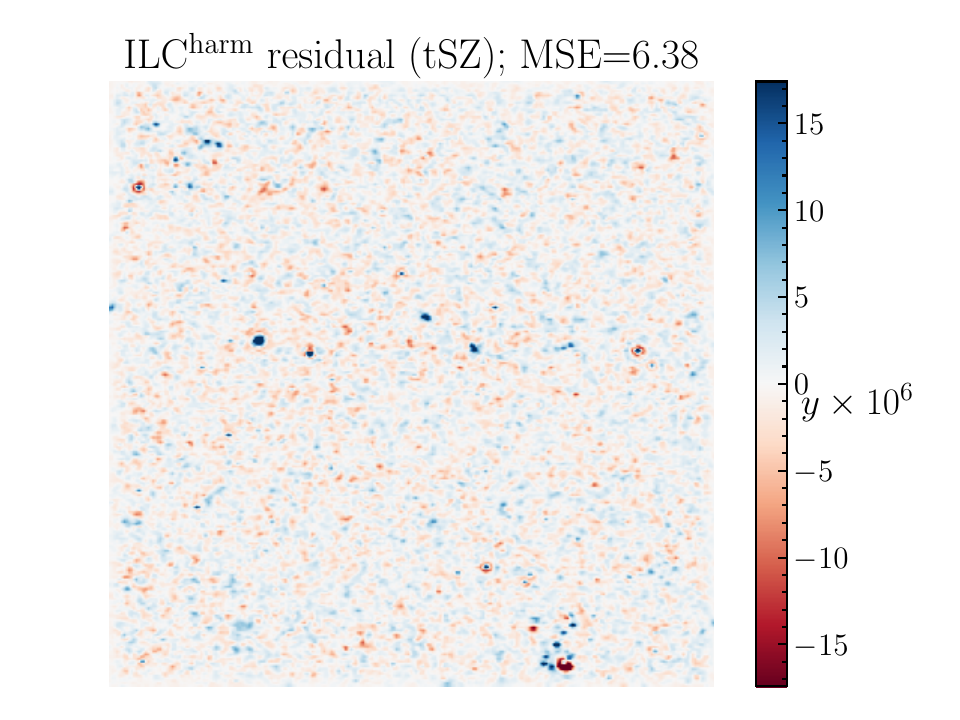}

\caption{A sample simulated multi-frequency small-scale mm-wave temperature measurement, from the Websky simulations. On the top row, we show the single-frequency maps; on the second row we show the true blackbody CMB+kSZ signal, along with the real-space ILC determination and the residuals of both the real-space and harmonic-space ILC estimates (defined by subtracting the truth from the ILCs). On the bottom row, we show the analagous plots for the $y$-map (tSZ) signal. {We show all plots without experimental noise, and the ILCs were performed here with no experimental noise.}.}\label{fig:sample_SO}
\end{figure}

\subsubsection*{Harmonic ILC}

For cases where there is a relevant scale evolution between the frequency dependences of the different components --- such as, for example, when there are different foreground components dominating at different scales, or cases where the foregrounds dominate on large scales and the instrumental noise on small scales --- it can be useful to perform ILC with different weights on different scales. In the isotropic case, this means that we can use the data to measure the covariance matrix as a function of multipole $\ell$:
\be
\hat C_{ij}(\ell) = \left<T^*_{\ell m;i} T_{\ell m;j}  \right>= \frac{1}{2\ell+1}\sum_{m}  T_{\ell m;i}^{*}{}T_{\ell m;j}{}
\ee
where $T_{\ell m;}{}_i$ are the spherical harmonic coefficients  of the measured temperature field $T_i(p)$. {In practice, $\ell$-bins of width $\Delta\ell>1$ can be used to estimate $\hat C_{ij}(\ell)$.}

\subsection{Performance of ILC and preservation of the signal}\label{sec:unbiased_ILC}

The ILC is always the linear combination of any maps that minimizes the variance of the resulting map of a component of interest.  For the \textit{Gaussian} case where the variance is the only statistic of interest, it is the optimal linear combination one can construct. 

Let us consider the performance of the ILC algorithm on simulations of the extragalactic temperature field at mm wavelengths, implemented on patches of side 3.8 degrees ($\sim 14.7$ square degrees) cut out from a whole-sky simulation. These are based on the Websky~\cite{2020JCAP...10..012S,2022JCAP...08..029L} simulations; they are described in detail in Appendix~\ref{app:simulations_extragal}. {They include a Gaussian primary CMB; a kSZ~\cite{1980MNRAS.190..413S} contribution; a tSZ~\cite{1969Ap&SS...4..301Z} contribution; a CIB contribution; and a contribution from radio point sources~\cite{2022JCAP...08..029L}. The patches have 256$\times$256 pixels, and have a Gaussian beam of full width at half maximum (FWHM) = 0.9 arcminutes applied.} Note that the kSZ effect preserves the CMB frequency spectrum and so any blackbody estimation is an estimation of CMB+kSZ anisotropies.

{It is important to note that the various signals involved in this scenario are correlated with each other, as they are sourced by objects which trace the same underlying cosmic web, and sometimes indeed by the same objects. However, we remove these correlations by rotating the CIB and radio maps such that they are no longer with the tSZ component in the patch to which they are assigned.}
 {In this section, we principally do this in order to in order to demonstrate the ``unbiased'' quality of the ILC.  We will consider the case with these correlations restored in Section~\ref{sec:realistic_fgs}}.  

{A sample patch is shown in Figure~\ref{fig:sample_SO}, where we include the full signal map at each frequency, along with the true blackbody (CMB+kSZ) and tSZ signals, their real-space ILC estimates, and the real- and harmonic-space ILC residuals (the error between the ILC estimations and the truth). By eye, it is clear that the ILC extracts the appropriate signal, even in these non-Gaussian cases. For the residuals, we indicate the mean-squared error (MSE), which quantifies the variance of the residual. }

\begin{figure}
    \includegraphics[width=0.49\textwidth]{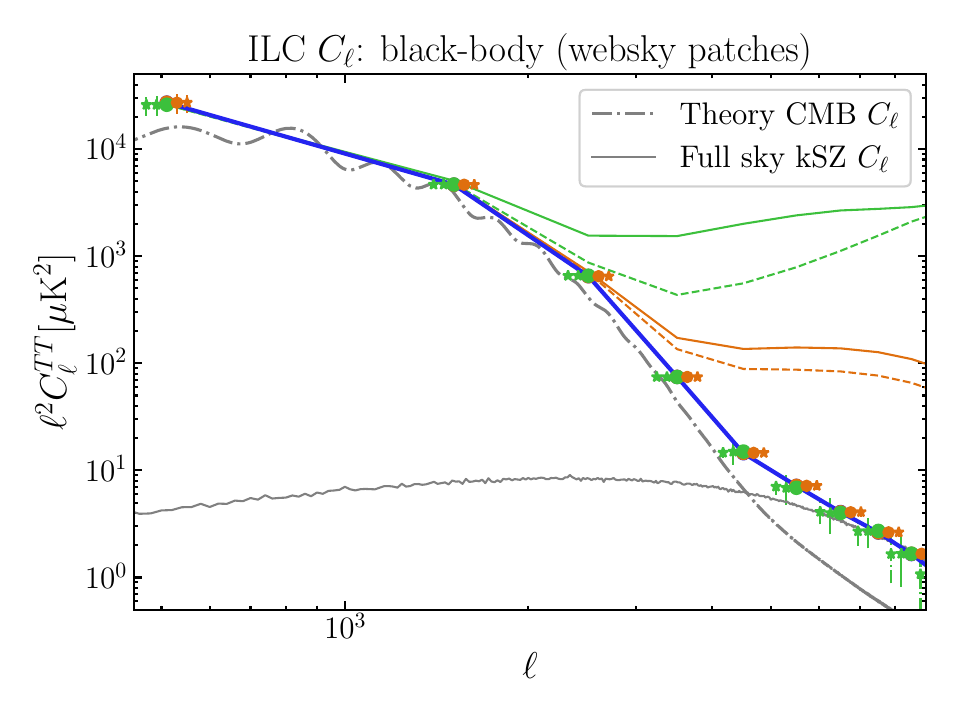}
    \includegraphics[width=0.49\textwidth]{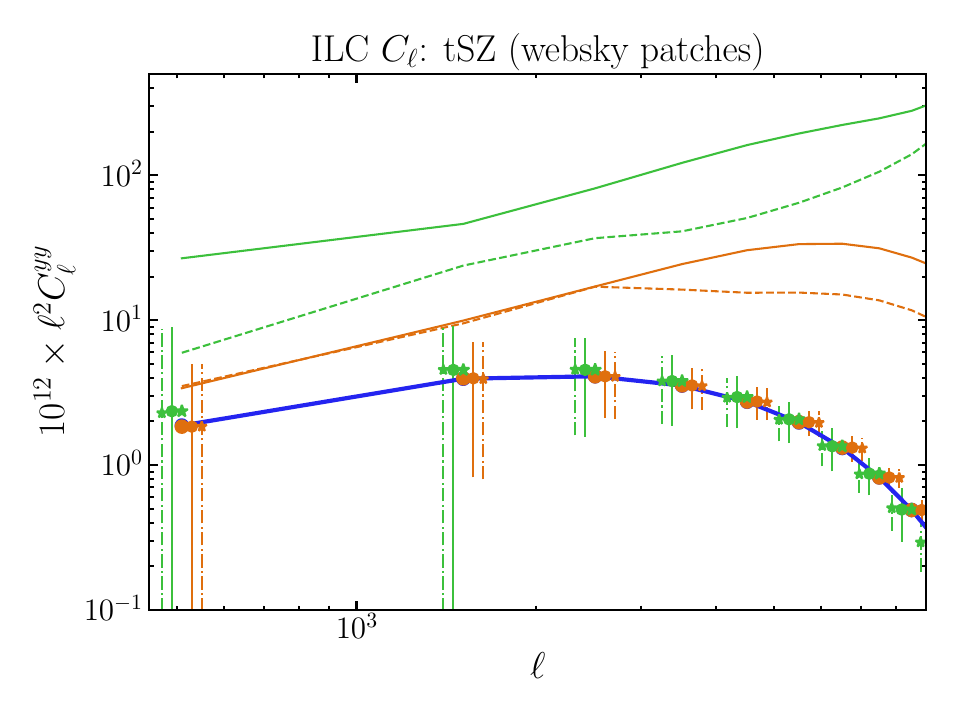}\\
    \includegraphics[width=0.55\textwidth]{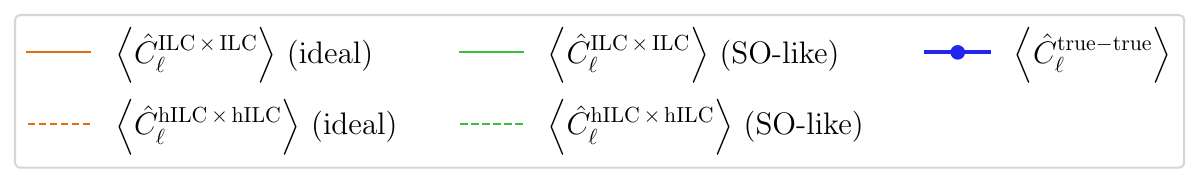}
    \includegraphics[width=0.44\textwidth]{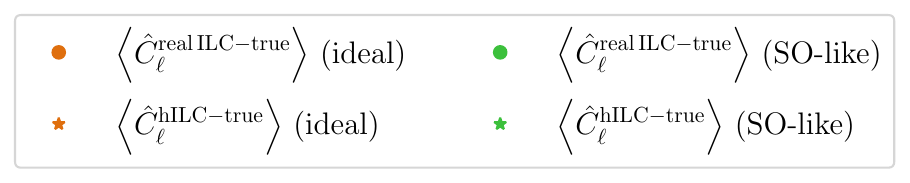}
\caption{ILC performance on the extragalactic simulations, both with no instrumental noise (orange lines, the ``ideal case''), and with Gaussian SO-like instrumental noise (green lines). We show the power spectra of the real-space and harmonic-space ILC estimations of both the blackbody component (CMB+kSZ, left) and the $y$-component (tSZ, right). In all cases we have performed the ILC separately on 128 $3\times 3$ square-degree patches, calculated the power spectra separately on the patches, and taken the mean (indicated by $\left<\cdot\right>$). We measure the auto-spectra of the true patches, the ILC estimations, and the cross-power spectra between the truth and the ILC estimations; these cross-spectra are indicated with the dots/stars for real space and harmonic space respectively. The error bars indicate the standard deviation of the power spectra measured on the patches. An important take-away from this plot is that the cross-spectra of the ILC estimations with the truth lie exactly on the $\left<\hat C_\ell^{\mathrm{true}-\mathrm{true}}\right>$ line: the ILC is an ``unbiased'' estimation of the truth. Note that all points are centered on the same values of $\ell$, but some are offset for clearer visualization.}\label{fig:power_true}
\end{figure}

\begin{table}
\begin{tabular}{|c||c|c|c|c|}\hline
Frequency [GHz] &93 &145 &225&280\\\hline\hline
Noise [$\mu$K-arcmin]   &5.8&6.3&15&37\\\hline
Beam FWHM (arcmin) &2.2&1.4&1.0&0.9\\\hline

\end{tabular}
\caption{The white noise levels and effective Gaussian beams applied to the SO-like simulations (see Table 1 of~\cite{2019JCAP...02..056A}).}\label{tab:SOnoise}
\end{table}
The auto-power spectra of the true and ILC signals are shown in Figure~\ref{fig:power_true}, for cases with no instrumental noise (as in the plots in Figure~\ref{fig:sample_SO}), and when we have added SO-like Gaussian white noise (see Table~\ref{tab:SOnoise}; here we consider only four of the planned six frequency channels of the SO wide-area survey, and we ignore the large-scale $1/f$ noise expected due to atmospheric effects).  We also show the cross-power spectra of the estimates with the truth. 
The cross-correlation between the estimate and the truth is exactly equal to the auto-power of the truth; this demonstrates the \textit{unbiased} nature of the ILC. {Note that for the tSZ case, this is an artifact of the fact that we have removed the correlations between the foreground CIB and radio components and the tSZ signal; in reality, the residual foregrounds will bias this measurement slightly due to their correlation with the true signal.}

\section{Signal-preserving ML predictions of a component}\label{sec:signal_preserving_vs_direct}
Inspired by the ILC, in this section we will introduce a framework to use ML to predict a signal of interest in a signal-preserving manner. In Section~\ref{sec:nonlinear_unbiased} we introduce the general framework, where we use signal-free inputs to a neural network to predict foregrounds; and in Section~\ref{sec:ILCcorrection} we will specialize the framework to the case where we predict the ILC residual and then subtract this from the ILC estimate.

It will be helpful to recall the underlying observation model:
\begin{equation}
T_i(p) = a_i T^{\mathrm{coi}}(p) + F_i(p) +N_i(p)
\end{equation}
where $i$ indicates a frequency channel and $p$ indicates the pixel; $ T^{\mathrm{coi}}(p)$ is the component one is interested in isolating; and $F_i(p)$ and $N_i(p)$ are the foregrounds and noise respectively, which are uncorrelated with $ T^{\mathrm{coi}}(p)$.
\subsection{Extension of the unbiased condition to non-linear combinations}\label{sec:nonlinear_unbiased}
ILC has been applied succesfully to many CMB datasets. However, many real-world signals are non-Gaussian and anisotropic. Such a situation is ideal for the application of an ML solution, and such solutions have been proposed in the literature (see, e.g.,~\cite{2020ApJ...903..104P,2022A&A...666A..89C,2023arXiv231007590C,2022ApJS..260...13W,2023arXiv230212378A,2022arXiv221202847B}).  One way of training an ML model to predict a component is to explicitly train on simulations to do so, given all the multifrequency measurements:
\be
\hat T^{\mathrm{pred},\,\rm{coi}} = f(T^i),
\ee
where $f(T^i)$ is some non-linear function that an ML model (such as a convolutional neural network (CNN)) can learn. We refer to such an approach as the ``direct-NN'' approach.  However, such a non-linear model loses the interpretability of the ILC, where the ILC condition ensures unbiasedness of the predicted or estimated component map. We thus impose such a condition on the function $f$, such that $f$ is of the form
\be
 f(T^i) = \sum _i w_iT_i + \tilde f(F, N),\label{linear_nonlinear}
\ee
where $w_i$ are weights that obey the ILC condition $\sum_i a_iw_i=1$,  
$\tilde f$ is another non-linear function, and $F$ and $N$ are the foregrounds and noise. This is equivalent to
\be
 f(T^i) = T^{\mathrm{coi}} +  \sum_i w_i (F_i+N_i) + \tilde f(F, N).
\ee
As long as $\tilde f$ does not contain information about the signal of interest {(and $F_i$ and $N_i$ are uncorrelated with the signal of interest)}, such a combination explicitly preserves the signal of interest in an unbiased manner.

We now propose to use an ML model to learn the non-linear function $\tilde f$. It can only take as inputs the foregrounds and noise.  We can construct signal-free inputs from the data as linear combinations
\be
\tilde f(F,N) \sim \tilde f\left(\left\{\sum_i c_i T_i\right\}\right)
\ee
where $\sum_i a_ic_i=0$, 
such that all the linear combinations that enter the ML model are signal-free.  We can then train a neural network to find $\tilde f$ such that some metric (loss function) is minimized. 

{One could make several choices for the signal-free inputs to $\tilde f$. Likewise there is freedom in the weights $w_i$ used to define $f$ (in Equation~\eqref{linear_nonlinear}). In our implementation, we choose these to be the weights that solve the minimum-variance condition (i.e., the ILC weights). For the signal-free linear combinations that enter $\tilde f$, we subtract the ILC estimation from the original frequency maps. We then train $\tilde f$ to predict the residual on the ILC maps, which we can then subtract from the ILC solution, resulting in a different estimate of $T^{\rm{coi}}$.}

\subsection{Estimating the ILC residual with signal-free maps}\label{sec:ILCcorrection}

Consider the ILC estimation of a component:
\be
\hat T^{\rm{ILC}}(p) = T^{\rm{coi}}(p) + \Delta T^{\rm{ILC}}(p),
\ee
where $\Delta T^{\rm{ILC}}(p)$ quantifies the ILC residual, i.e., its difference from the true signal.

We wish to estimate  $\Delta T^{\rm{ILC}}(p)$ 
from the data, and subtract it from the ILC. If our estimation $\hat{\Delta T^{\rm{ILC}}}(p)$ is better (by some metric) than the original estimation $\hat T^{\rm{ILC}}$ (i.e., if $||\hat{\Delta T^{\rm{ILC}}}(p) - \Delta T^{\rm ILC}(p)||$ is smaller than $|| \hat T^{\rm{ILC}}-T^{\rm{coi}}||$  where $||\cdot||$ is some distance measure), we can then improve the component-of-interest estimation (i.e., reduce $||\hat T^{\rm{coi}}-T^{\rm{coi}}||$) by subtracting $\hat{\Delta T^{\rm{ILC}}}$ from $ T^{\rm{ILC}}$.

While, in principle, we can use \textit{any} linear combination of maps that obey this condition to estimate $\hat{\Delta T^{\rm{ILC}}}$, in practice we use the combination defined by
\be
\hat F_i \equiv T_i -  \hat T^{\rm{ILC}} \,, \label{foreground_estimation}
\ee
i.e., we calculate the ILC estimate of the component of interest and subtract it from each frequency map. As a reminder, if we replaced $\hat  T^{\rm{ILC}} $ in Equation~\eqref{foreground_estimation} with the true map of the component of interest, we would be left with exactly the foregrounds $F_i$ (and noise) in channel $i$.  It is for this reason that we refer to the quantity in Equation~\eqref{foreground_estimation} as $\hat F_i$, as it is some estimation of the foregrounds in channel $i$. Then, using simulations for which we know the true signal, we train a neural network, which takes as input $\hat F_i$, to predict the ILC residual $\hat{\Delta T^{\rm{ILC}}}$.  Finally, we subtract this quantity from the ILC map to build an ``improved'' ILC estimate of the component of interest. Thus our final prediction for the map of the component of interest is
\be
\hat T = T^{\rm{ILC}} - \hat{\Delta T^{\rm ILC}}.
\ee
Because $\hat{\Delta T^{\rm{ILC}}}$ is explicitly not dependent on the signal, the new $\hat T$ retains many of the desirable properties of the ILC. In particular, it is explicitly unbiased with respect to the component of interest, in that (for cases when the foregrounds and the signal are uncorrelated), it obeys
\be
\left< \hat T T \right> = \left< TT \right>
\ee
by design.

\section{Demonstration: small-scale extragalactic temperature}\label{sec:small_scale_results}

In this section, we describe the results of the network trained on high-resolution extragalactic temperature simulations. In Section~\ref{sec:simtrainingt} we describe the training set and the training procedure. In Section~\ref{sec:websky_on_websky} we demonstrate the results of the trained network when applied to simulation patches outside of the training set but made from the same simulation that they were trained on. In Section~\ref{sec:websky_on_sehgal} we demonstrate the results of the trained network when applied to simulation patches created with a different simulation. 

In Section~\ref{sec:real_harmonic} we comment briefly on the improvement with respect to real-space and harmonic ILC, and in Section~\ref{sec:realistic_fgs} we demonstrate the performance on the more realistic case where we have restored the correlations between the foregrounds and tSZ component.

\subsection{Simulations and training}\label{sec:simtrainingt}
\begin{figure}[h!]

\includegraphics[width=0.24\textwidth]{256pixels_ILCresidualCMB_plus_kSZ_smoothed.pdf}
\includegraphics[width=0.24\textwidth]{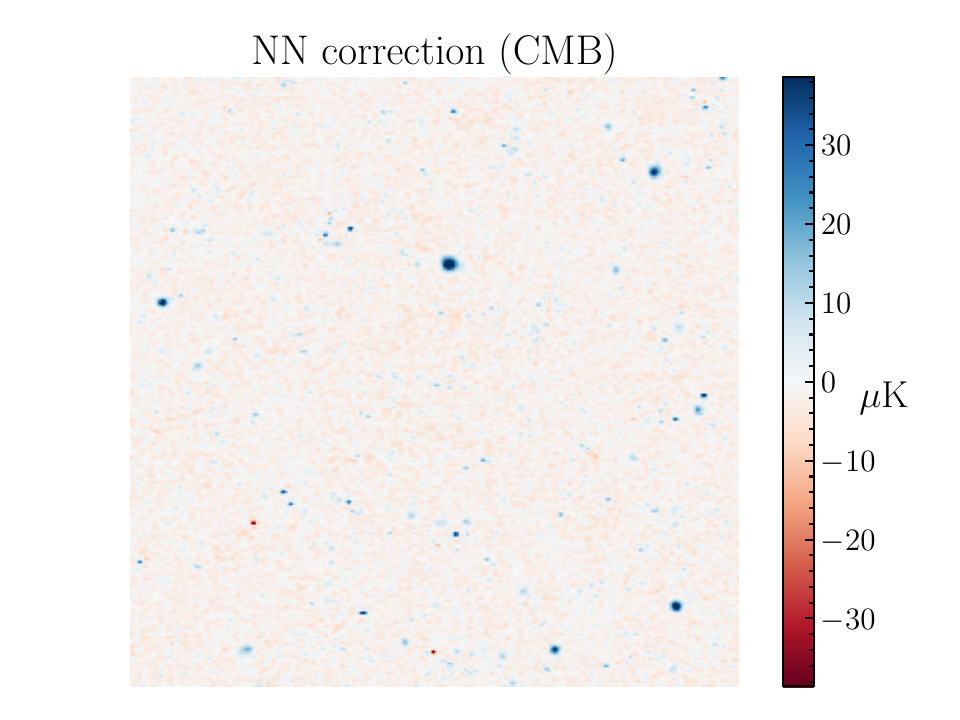}
\includegraphics[width=0.24\textwidth]{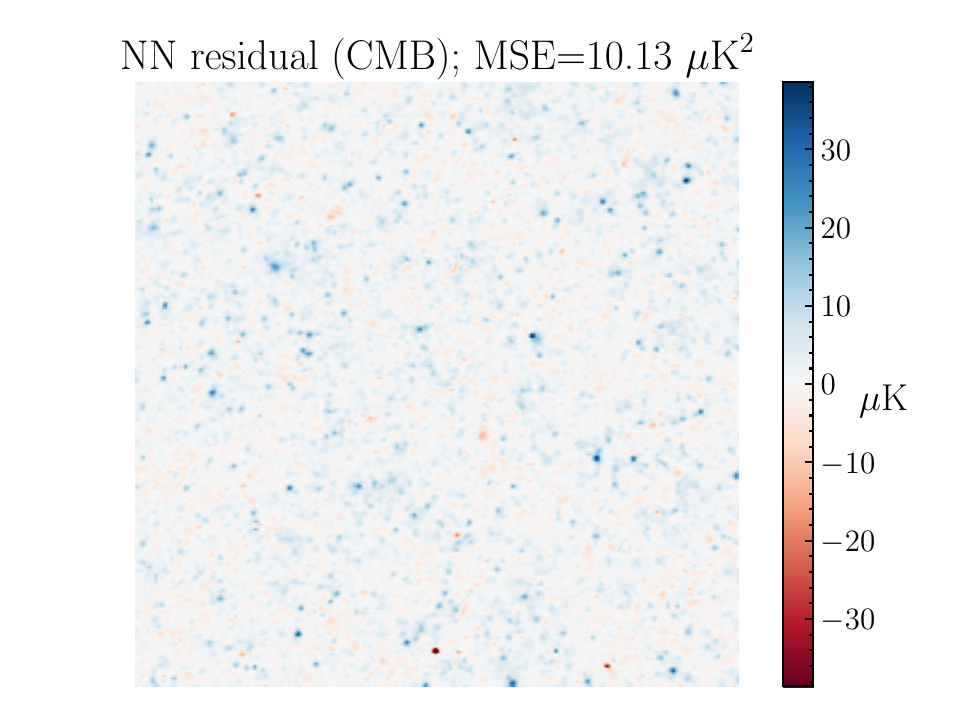}
\includegraphics[width=0.24\textwidth]{256pixels_NNharmresidualCMB_plus_kSZ_smoothed.pdf}

\includegraphics[width=0.24\textwidth]{256pixels_ILCresidualtsz.pdf}
\includegraphics[width=0.24\textwidth]{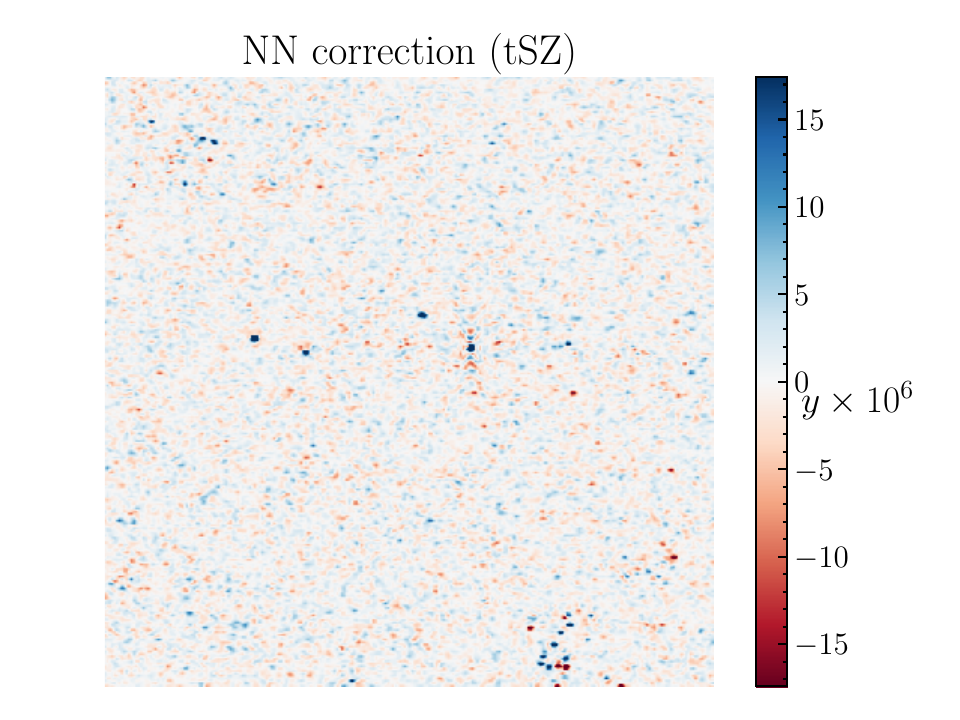}
\includegraphics[width=0.24\textwidth]{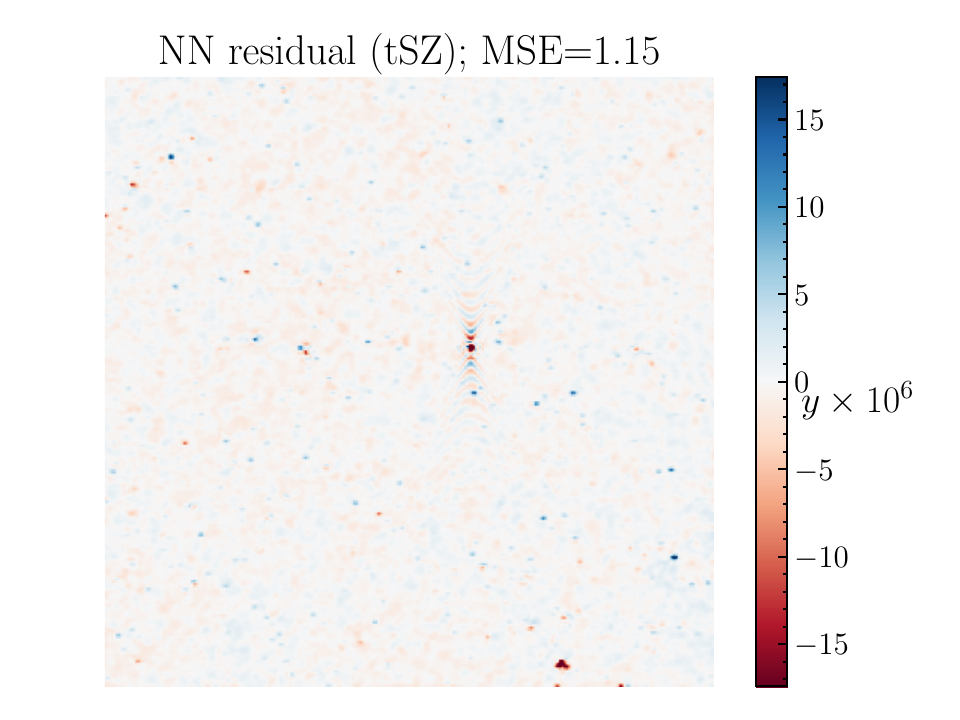}
\includegraphics[width=0.24\textwidth]{256pixels_NNharmresidualtsz.pdf}

\caption{The results of applying the NN to the patch shown in Figure~\ref{fig:sample_SO}. We train on the residual of the true ILC (which we define as ILC-truth), and the output of the NN (``NN correction'') is then subtracted from the ILC solution to get the NN prediction for the signal. The residual of the NN prediction is shown in the third column; we see that it is significantly lower than the real-space ILC residual. However, a harmonic ILC can also remove a significant amount of this foreground power, so we also include a harmonic ILC residual for comparison; we see the NN solution has succesfully removed a significant amount of point sources compared to this, and has lower mean-squared error. In this plot, we show only the idealized case without instrumental noise.}\label{fig:examples_nonoise_patches_extragalT}
\end{figure}

\begin{figure}[t!]
\includegraphics[width=0.49\textwidth]{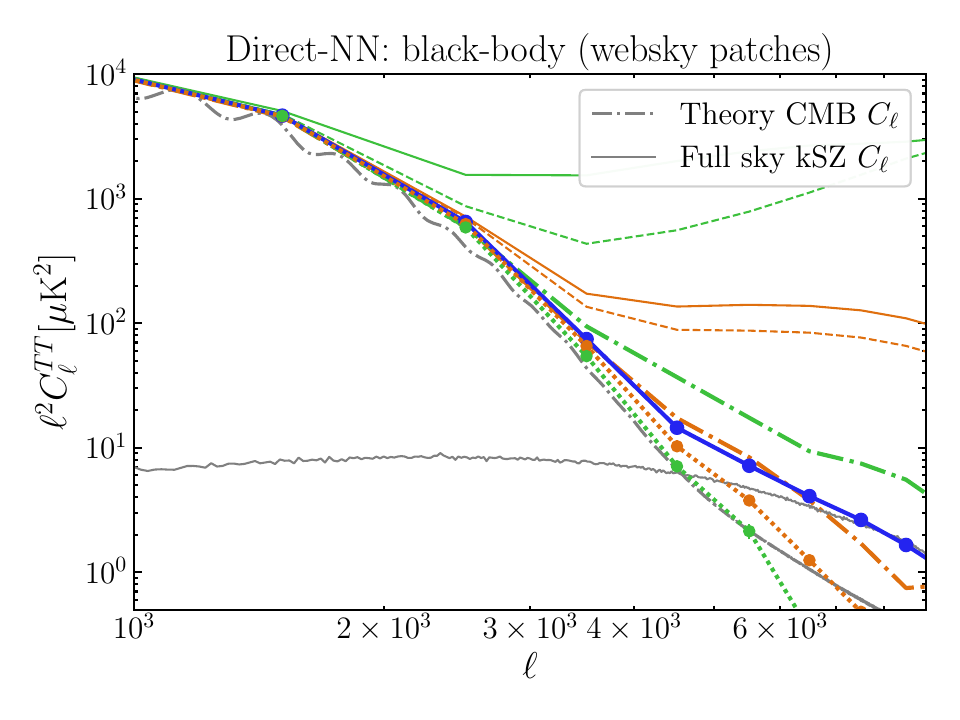}
\includegraphics[width=0.49\textwidth]{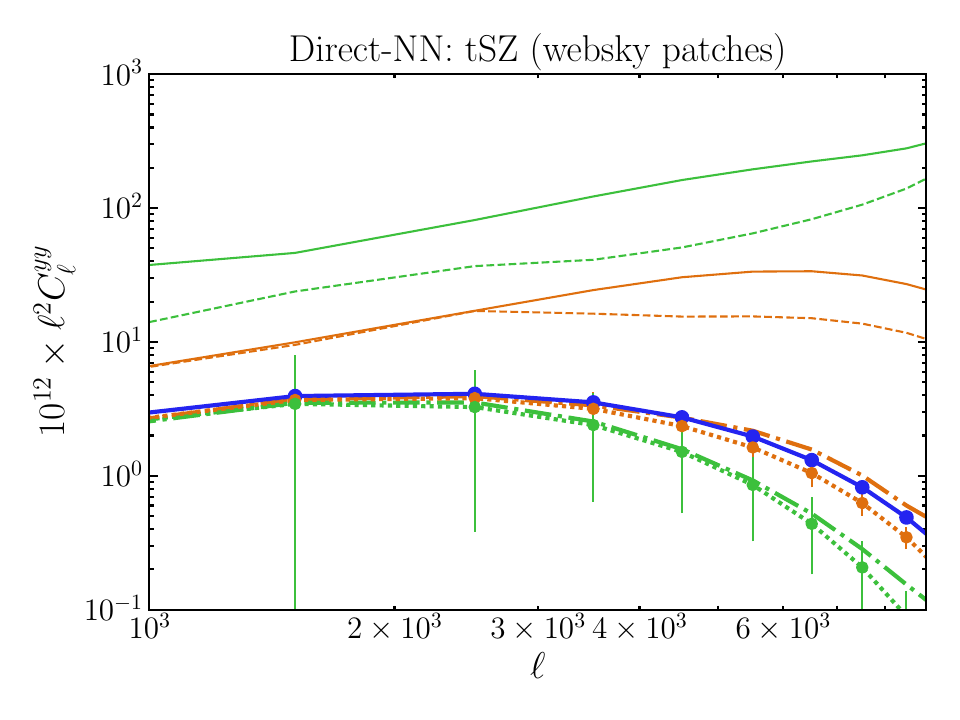}
\includegraphics[width=0.49\textwidth]{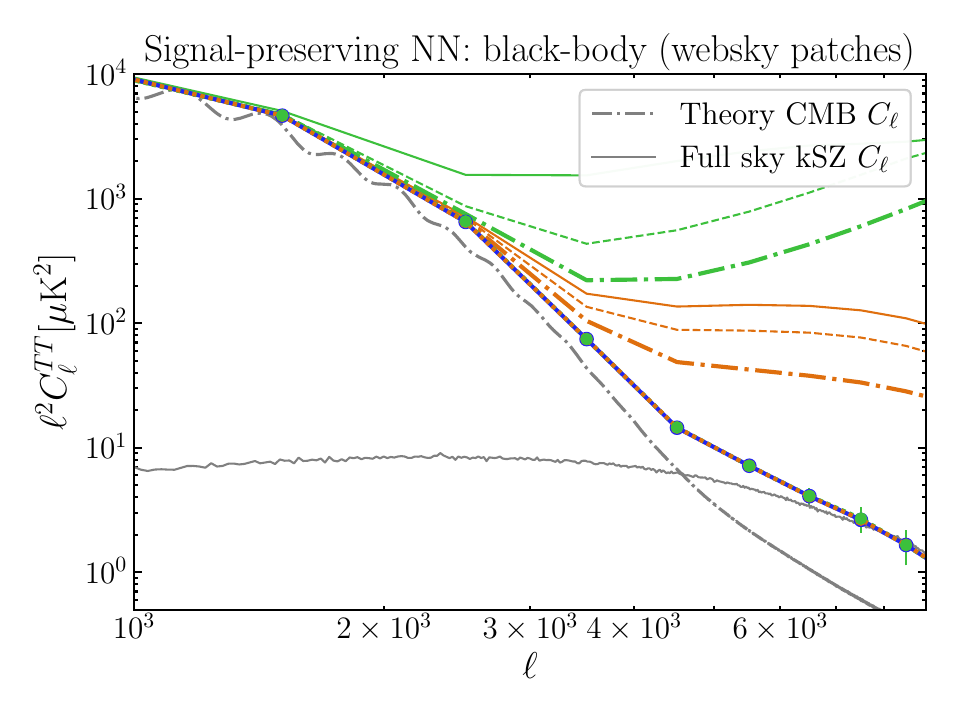}
\includegraphics[width=0.49\textwidth]{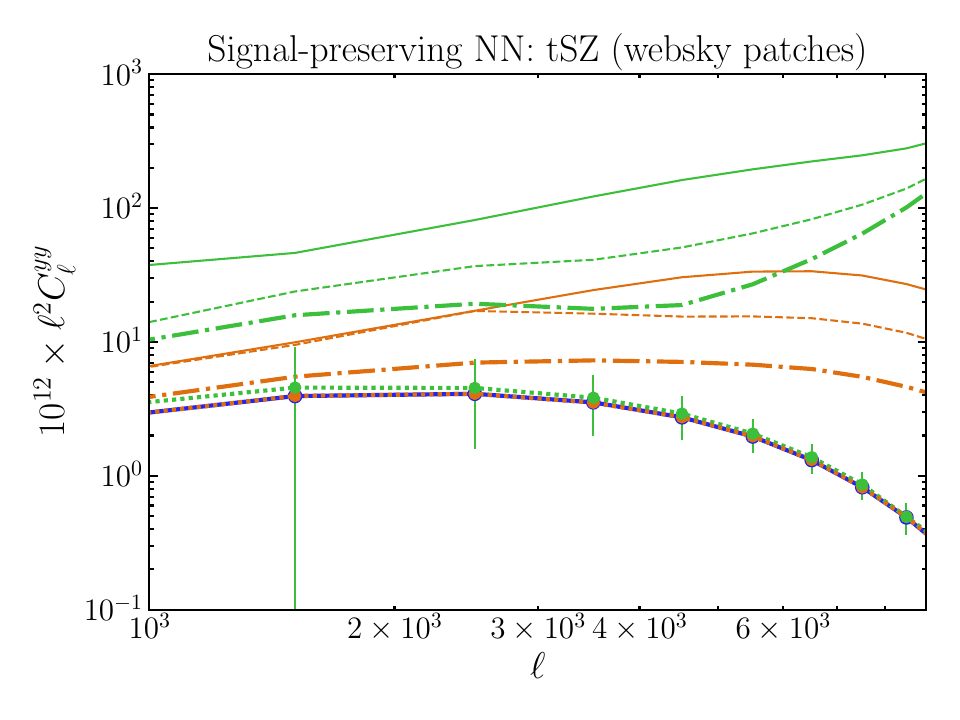}
\includegraphics[width=0.6\textwidth]{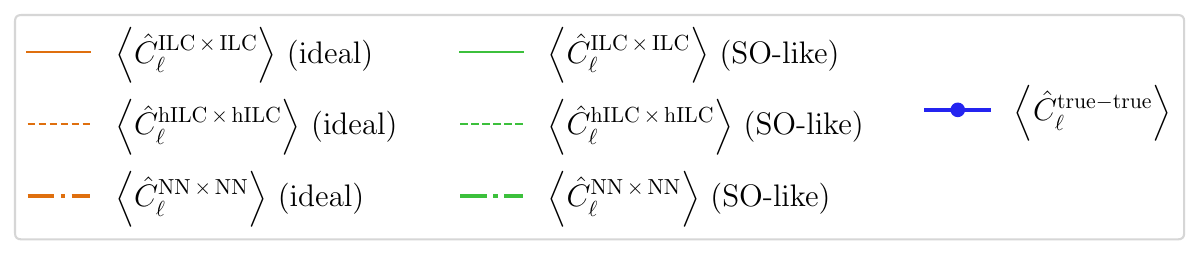}
\includegraphics[width=0.2\textwidth]{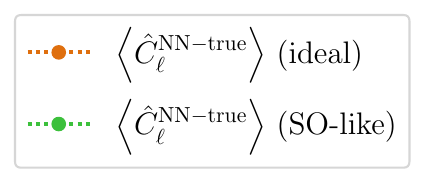}
\caption{Neural-network recovery of the blackbody (left) and Compton-$y$ (right) components on the Websky simulations, with and without SO-like noise. In the top two plots we show the direct prediction, where we include all of the data as an input to the ML model, and predict the blackbody or tSZ component. In the bottom two plots we show the signal-preserving version, where we show signal-free linear combinations to the ML model and predict the ILC residual. We include on each plot the real-space ILC and the harmonic-space ILC power spectra. In all cases we plot the mean power of 128 patches. We also show the true power spectrum of the relevant components (blue), and the cross-spectra of the predictions with the truth. While the naive predictions achieve lower power, they have a poor cross-correlation with the truth, indicating that they are highly biased. In contrast, the $\Delta T^{\rm{ILC}}$ predictions manifestly conserve the signal, as indicated, while also achieving significantly lower power than the ILC estimations. These plots are all shown for Websky patches that were not in the training set, although the network was trained on Websky patches.  }
\label{fig:ksz_websky}
\end{figure}

 Our training simulations are the same ones used in Section~\ref{sec:ILC} to demonstrate the results of ILC. They are based on the Websky~\cite{2020JCAP...10..012S} simulations; we describe them in detail in Appendix~\ref{app:simulations_extragal}. 
We train on square patches of side 3.8 degrees ($\sim$14.7 square degrees) cut out from a whole-sky simulation {as in Ref.~\cite{2021PhRvD.104l3521H}}.  The patches have square pixels of side-length 0.9 arcmin (thus they have $256\times256$ pixels in total). A sample patch is shown in Figure~\ref{fig:sample_SO}, including the single-frequency maps, the true CMB+kSZ and tSZ maps, and the ILC estimations of both CMB+kSZ and tSZ components along with the ILC residuals.

There are $\approx 3000$ {(non-overlapping)} patches in total, cut out from the whole-sky simulation. 
We use 1024 random patches in our training set, and validate on 128 random patches that are not used in 
the training set.
In all cases we train using a batch size of 32 with the loss function
\be
L=\sum_{p}\left(\hat {\Delta T^{\rm{ILC}}}(p)-\Delta T^{\rm{ILC}}(p)\right)^2.
\ee
 During training, we keep track of both the training loss and a validation loss, measured on the training and validation patches respectively; we stop training our networks when the validation loss stops decreasing.

\textit{After training}, we also apply our network, trained on Websky simulations, to 128 analogous patches extracted from the Sehgal et al.~\cite{2010ApJ...709..920S} simulations. We do this to illustrate the \textit{proof of concept} that our method can be applied to simulations on which the neural network was not trained, while retaining some desirable properties, e.g., that the result is unbiased with respect to the true signal. We stress that we have not fully optimized our method (e.g., extensively tuned hyperparameters or optimized the ML model to the problem), as we wish to generically illustrate the application of our method. In reality, it would be ideal to to add more variation to the training set, such as through the inclusion of a broader array of simulations, before applying to data.

In the training set, we always use patches in which there are no correlations between the foregrounds and the signal of interest. This is not a true approximation of reality. The Websky simulations include such foregrounds, as the components all trace the same underlying large-scale-structure distribution; we remove the correlations by assigning CIB and radio patches to tSZ patches from a different place in the sky (i.e., rotating the maps before projecting to patches). In Section~\ref{sec:unbiased_ILC} we did this in order to clearly demonstrate the unbiased nature of the ILC. While realistic correlations can be restored by associating the CIB and radio patches with the correct tSZ patches, we \textit{never do this in the training set}, as we do not want the ML model to learn about the true signal through its correlation with the residual foregrounds that leak through. These correlations can be restored at implementation time, after training, as we will do in Section~\ref{sec:realistic_fgs}.

In every case, we use four frequency channels relevant to the Simons Observatory: 93, 145, 220, and 280 GHz (as the Sehgal et al.~simulations are not provided at these frequencies, we use their 143 and 217 GHz CIB and radio point source maps as proxies for the 145 and 220 GHz emission). We assume perfect calibration and delta-function passbands.  We separately train networks to isolate the blackbody CMB+kSZ component and the tSZ component in the maps, following the procedure described above. In each case, we consider separately an idealized case with no instrumental noise and 0.9$^\prime$ beams, and a case with SO-like Gaussian instrumental noise and beams. As the ILC is optimal for the Gaussian case, we expect less improvement over the ILC in the case where we include Gaussian noise. 

In Figure~\ref{fig:examples_nonoise_patches_extragalT} we show explicitly the predictions of the neural network when applied to the patch in Figure~\ref{fig:sample_SO}, in the idealized no-instrumental-noise case, both when trained to predict the corrections to the ILC for the blackbody (CMB+kSZ) and the tSZ components. It is clear that the NN can predict non-Gaussian features, such as the extended tSZ sources and the point sources that are present as residuals in the blackbody case. Below, we quantify in more detail the comparison between the NN predictions and the ILC estimates.

In Section~\ref{sec:websky_on_websky} we present the results when we apply the Websky-trained NNs to the Websky patches in the validation set. In Section~\ref{sec:websky_on_sehgal} we present the results when apply the Websky-trained NNs to the patches from the Sehgal et al.~simulations.

\subsection{Performance on the Websky simulations} \label{sec:websky_on_websky}

In Figure~\ref{fig:ksz_websky}, we show some results when we train separately to recover both the blackbody (CMB+kSZ) component and the Compton-$y$ component, with and without instrumental noise, for the Websky simulations. In the top two plots we compare to a case where we have directly trained a neural network to predict the relevant component using all of the frequency channels as information (i.e., we are not training to learn the corrections to the ILC in these plots). In this case, we use the loss function
\be
L^{\mathrm{direct}}=\sum_p \left( \hat T(p) - T(p)\right)^2.
\ee
We refer to this method of training as the direct-NN case, in contrast to the signal-free method of training which we refer to as signal-free-NN.

In the bottom two plots we have trained a network to predict the ILC residual ${\Delta T^{\rm ILC}}$, using only signal-free linear combinations as input, and subtracted this from the ILC. In all cases we include the analogous results for the real-space and harmonic-space ILC (HILC), in solid and dashed lines respectively, with the auto-power spectra of the NN results in dot-dashed lines. The cross-power spectra of the NN predictions with the truth are shown in dotted lines. In the direct-NN case, while we get a lower-noise estimation of the true power spectrum (blue), the  realizations do not have a good correlation coefficient with the truth, indicating that they are highly biased. In the signal-free-NN case, the NNs outperform the ILCs, and retain the unbiased property that $\left< \hat T^{\mathrm{pred}} T^{\mathrm{true}}\right> = \left<  T^{\mathrm{true}} T^{\mathrm{true}}\right>$.

\begin{figure}[t!]
\includegraphics[width=0.49\textwidth]{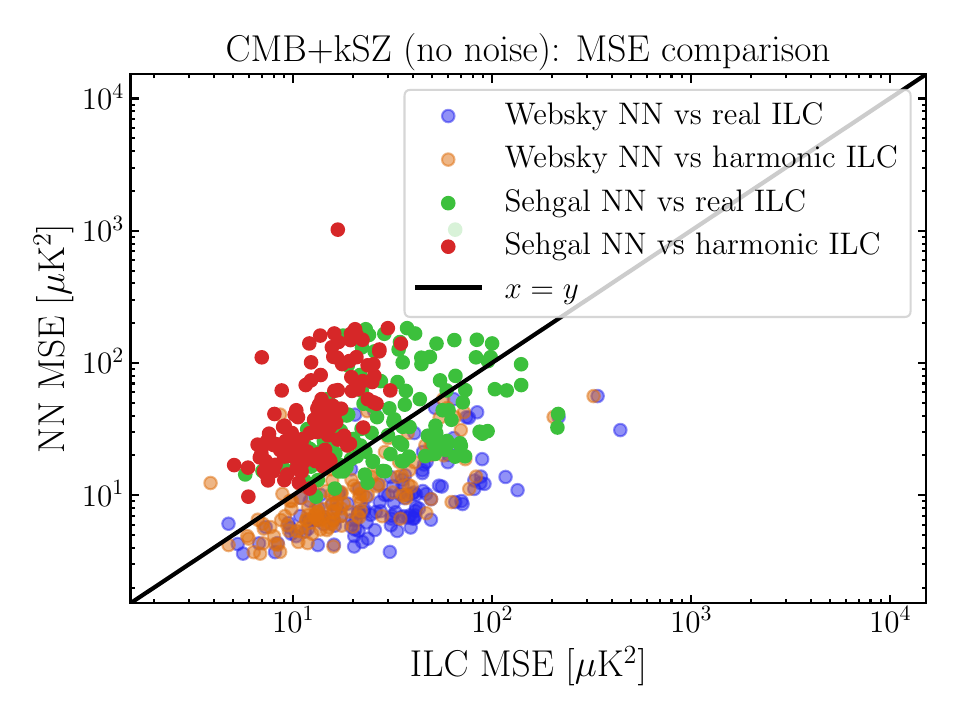}
\includegraphics[width=0.49\textwidth]{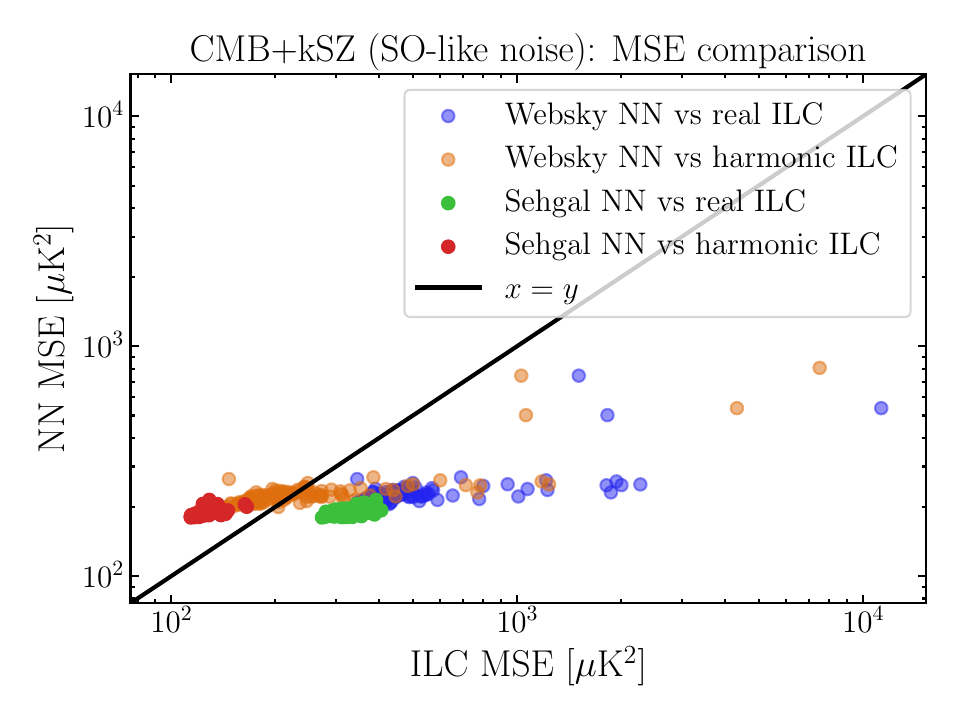}
\caption{The per-patch mean-squared-error (MSE) of the predicted blackbody components, plotted against the per-patch MSE of the ILC prediction of the same patch. We consider the no-noise case on the left, and the SO-like-noise case on the right. In every case we plot the line $x=y$; a point below this line indicates that the prediction has a lower MSE than the corresponding ILC prediction. We show this both for the real-space ILC (which is our starting point) and the harmonic ILC, and for both the Websky and the Sehgal et al.~simulations. }\label{fig:MSE_blackbody}
\end{figure}

\begin{figure}[h!]
\includegraphics[width=0.49\textwidth]{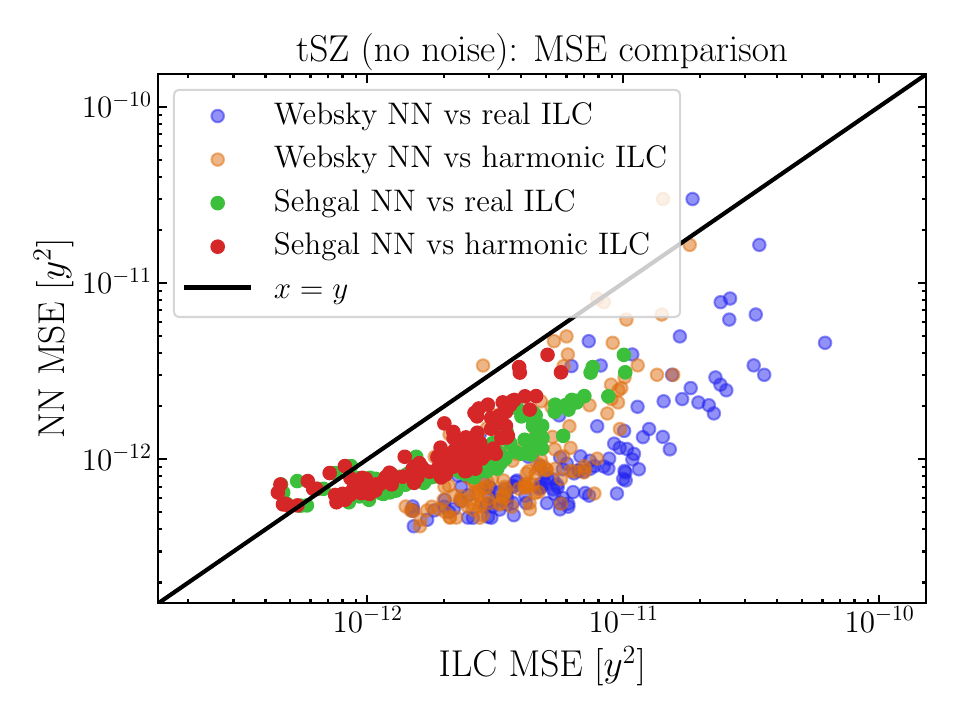}
\includegraphics[width=0.49\textwidth]{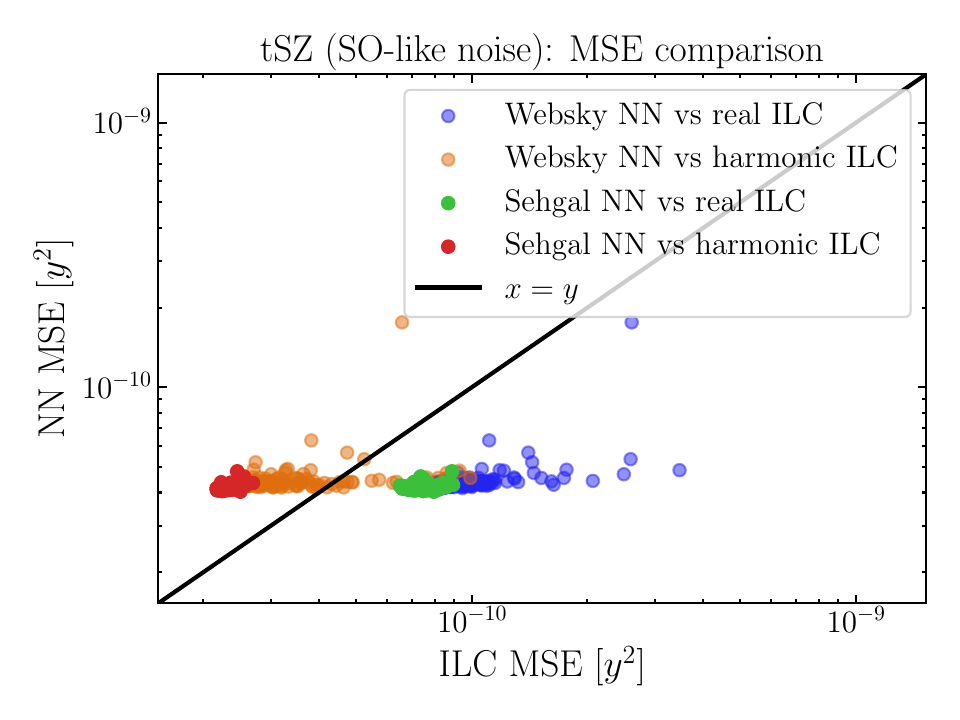}
\caption{The per-patch MSE of the predicted tSZ map, plotted against the per-patch MSE of the ILC estimate of the same patch. We consider the no-noise case on the left, and the SO-like-noise case on the right. In every case we plot the line $x=y$; a point below this line indicates that the prediction has a lower MSE than the corresponding ILC prediction. We show this both for the real-space ILC (which is our starting point) and the harmonic ILC, and for both the Websky and the Sehgal et al.~simulations. }\label{fig:MSE_tSZ}
\end{figure}

We plot the per-patch MSE of the final predicted components against the ILC MSE in Figures~\ref{fig:MSE_blackbody}  (blackbody component) and~\ref{fig:MSE_tSZ} ($y$ component), with the no-noise case on the left and the SO-like-noise case on the right. On these plots, we indicate the line $x=y$; thus a point underneath (to the right of) this line indicates that the NN is outperforming the relevant ILC by this metric. By this metric, the network always outperforms real-space ILC, and also always outperforms  HILC in the no-noise case.

\subsection{Performance on the Sehgal et al.~simulations} \label{sec:websky_on_sehgal}

\begin{figure}[t!]\includegraphics[width=0.49\textwidth]{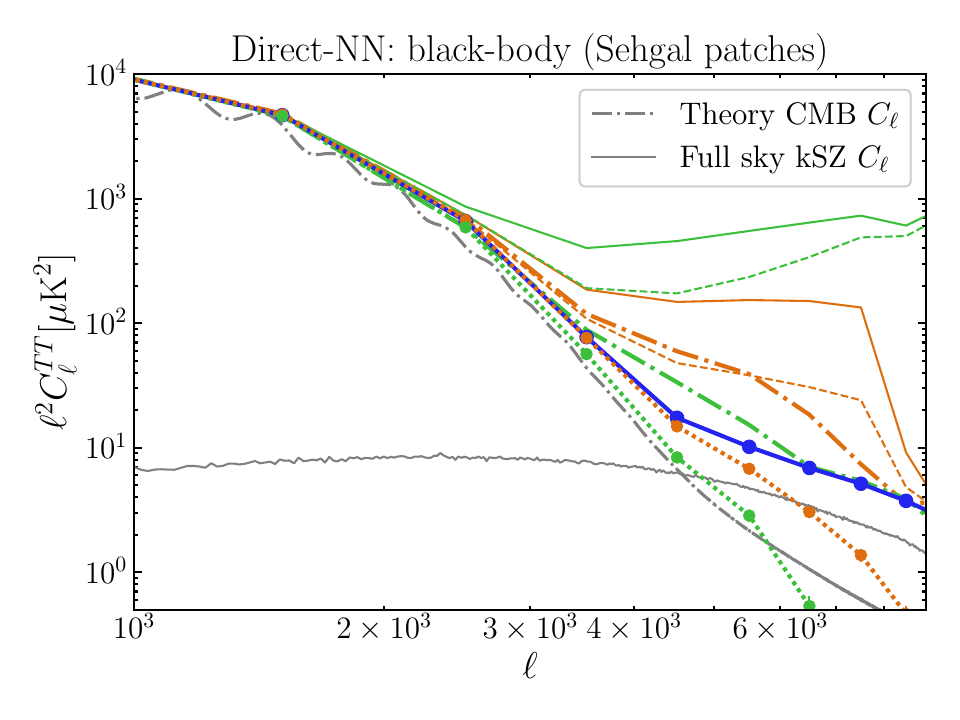}
\includegraphics[width=0.49\textwidth]{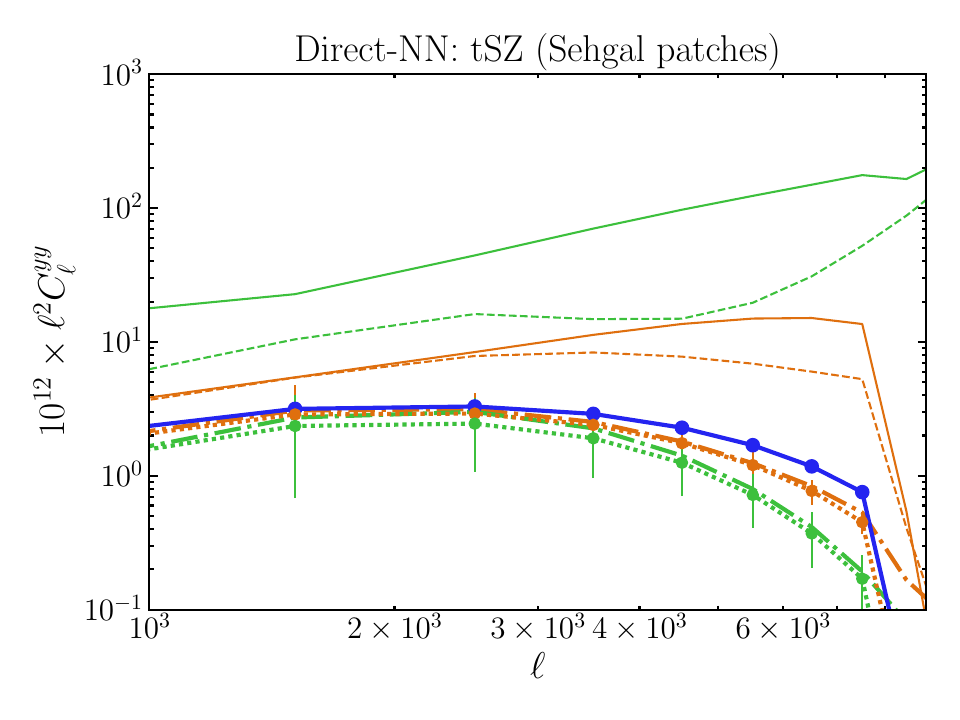}
\includegraphics[width=0.49\textwidth]{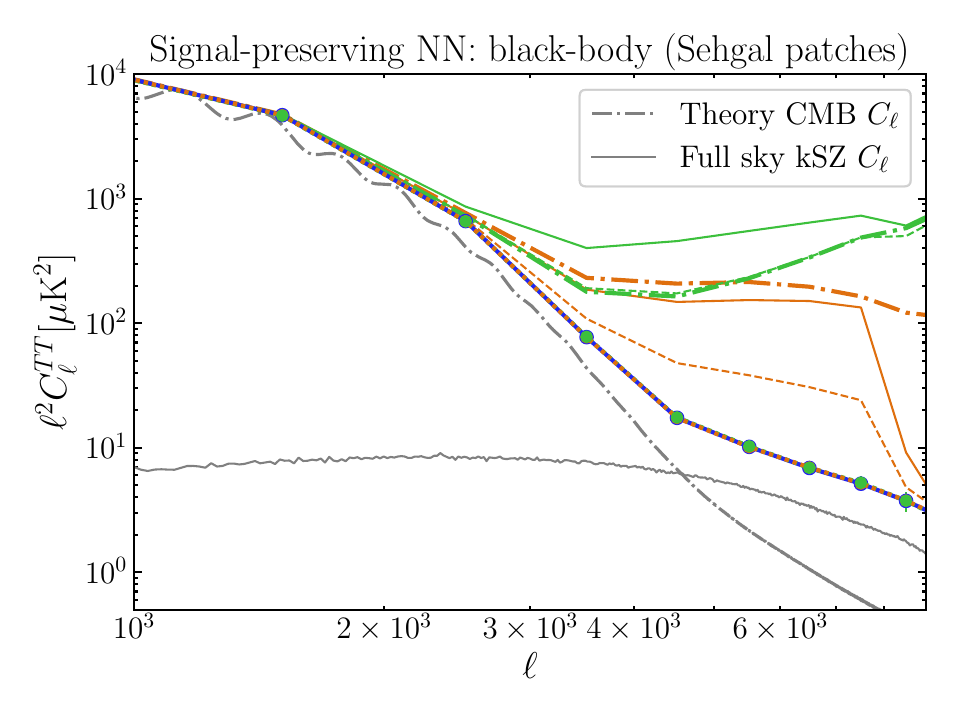}
\includegraphics[width=0.49\textwidth]{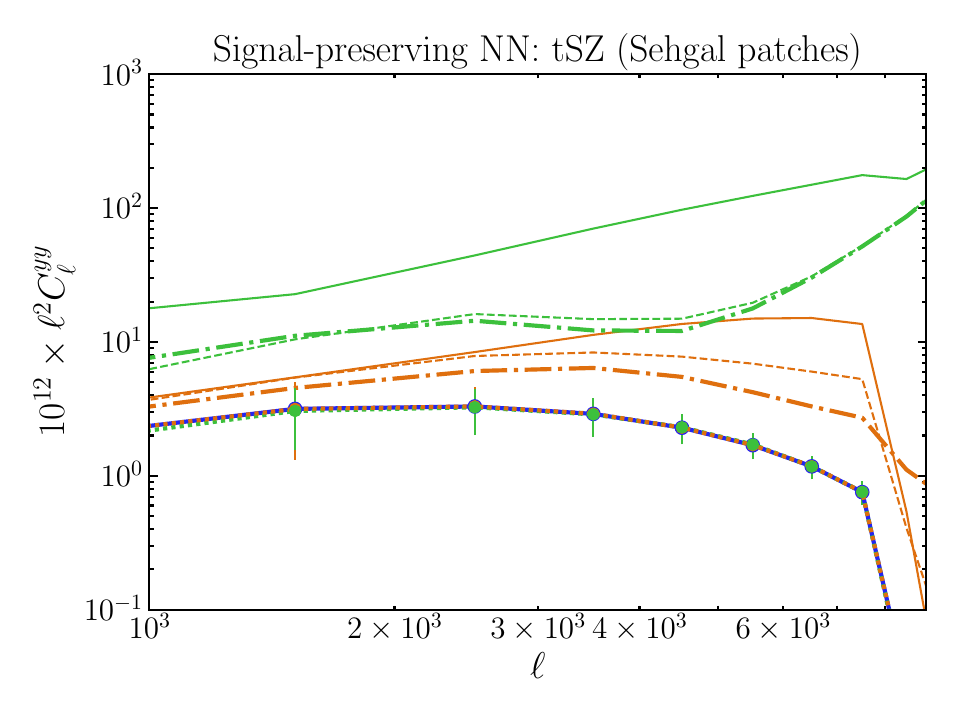}
\includegraphics[width=0.6\textwidth]{leg_autoCLsILC_withNN.pdf}
\includegraphics[width=0.2\textwidth]{leg_Xcrosstrue_NN.pdf}
\caption{As for Figure~\ref{fig:ksz_websky}, but for the Sehgal et al.~simulations (analyzed using our NNs trained on the Websky simulations). We again see improvement over the original ILC (orange), and unbiasedness compared to the naively-trained network (red). }\label{fig:ksz_sehgal}
\end{figure}

We show in Figure~\ref{fig:ksz_sehgal} the results of the networks trained on Websky when applied to patches extracted from the Sehgal et al.~extragalactic simulations. Again, the power spectra are measured from the mean of 128 patches. The distribution of the per-patch MSEs is also included in Figures~\ref{fig:MSE_blackbody} and~\ref{fig:MSE_tSZ}. We stress that we have not expected our networks to perform well on these patches, as the NNs were trained on entirely different simulations.  However, we wish to indicate the preservation of the unbiased constraint (as indicated by the dotted lines in Figure~\ref{fig:ksz_sehgal}): $\left<C_\ell^{\mathrm{NN}-\mathrm{true}}\right>=\left<C_\ell^{\mathrm{true}-\mathrm{true}}\right>$ for the signal-free-NN case. Interestingly, we do still outperform ILC in some cases, especially in the no-noise tSZ case where the network even tends to outperform the HILC.

\subsection{Performance compared to real-space and harmonic-space ILC}\label{sec:real_harmonic}

{In general, the networks tend to perform better than real-space ILC, which is encouraging as the real-space ILC is always their starting point. It would certainly be interesting to experiment further with the loss function to try to outperform harmonic-space ILC, either by perhaps starting from the harmonic-space ILC estimate, or by upweighting different scales in the loss function; we leave such exploration to future work. }

\subsection{Including realistic foreground correlations}\label{sec:realistic_fgs}

To illustrate the proof-of-concept of our method, we have removed the correlations between the foregrounds and the signal, which in practice cause some bias between the recovered signal and the foregrounds due to the correlation of the residual foregrounds with the truth. However, in truth these correlations exist.

\begin{figure}
\includegraphics[width=0.49\textwidth]{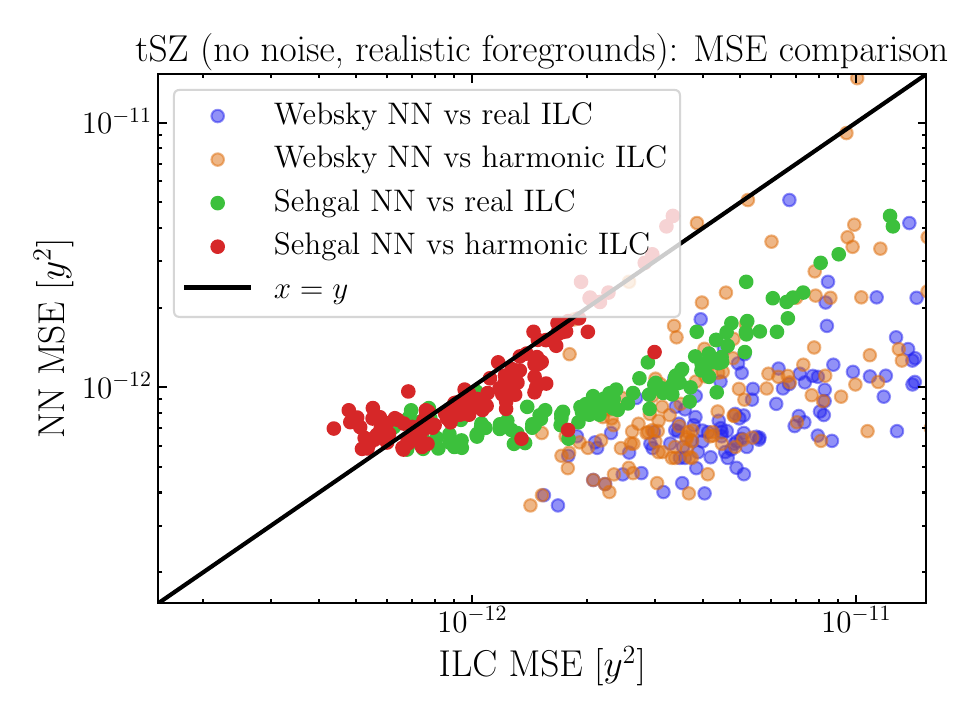}
\includegraphics[width=0.49\textwidth]{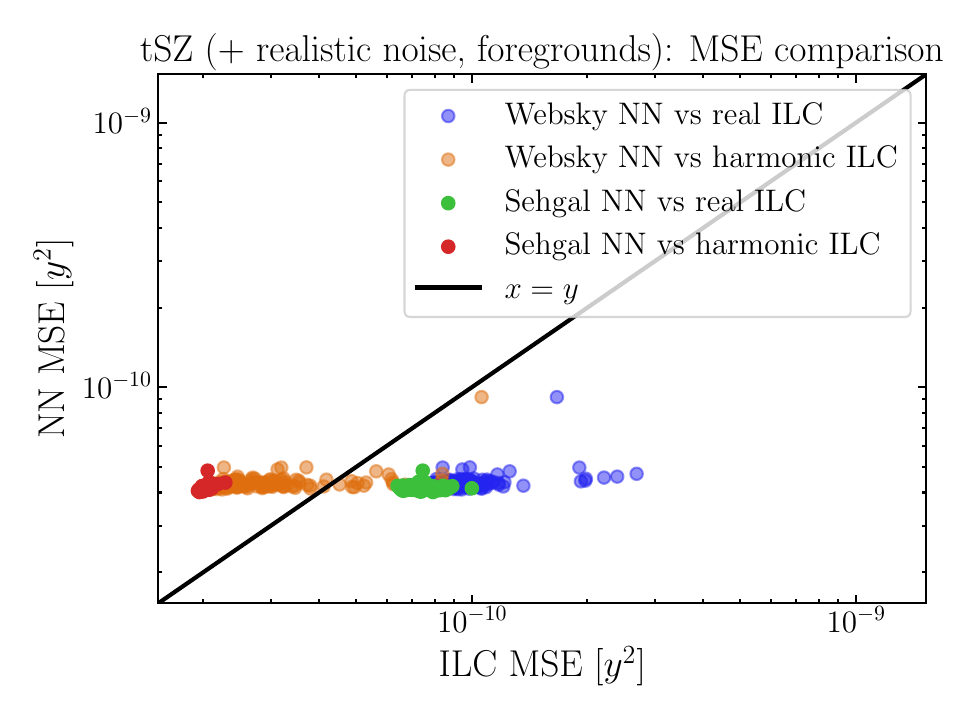}
\includegraphics[width=0.49\textwidth]{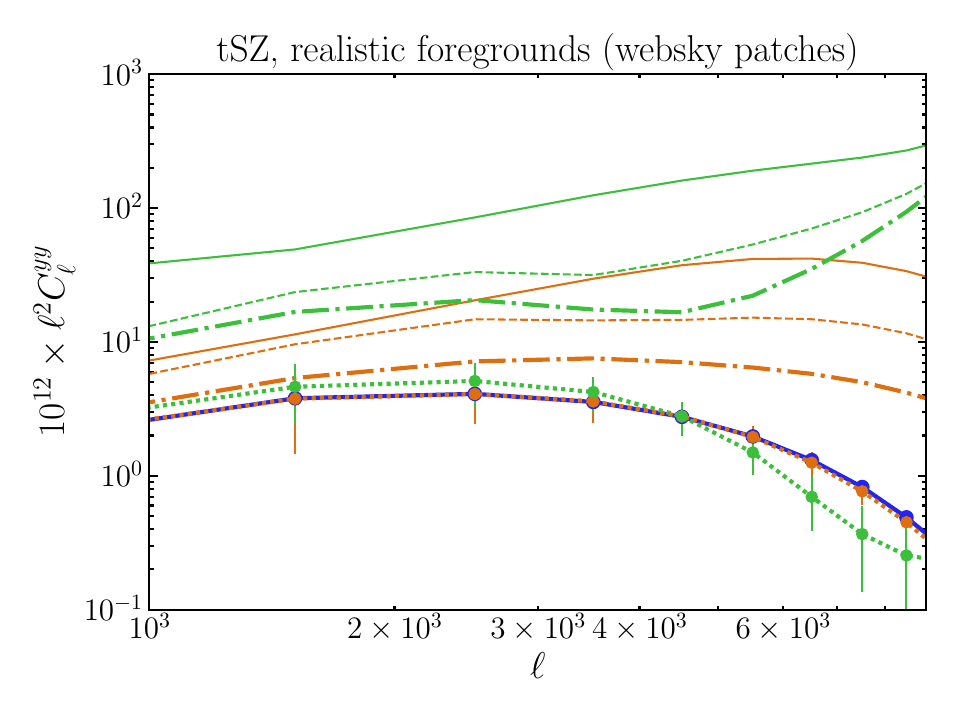}
\includegraphics[width=0.49\textwidth]{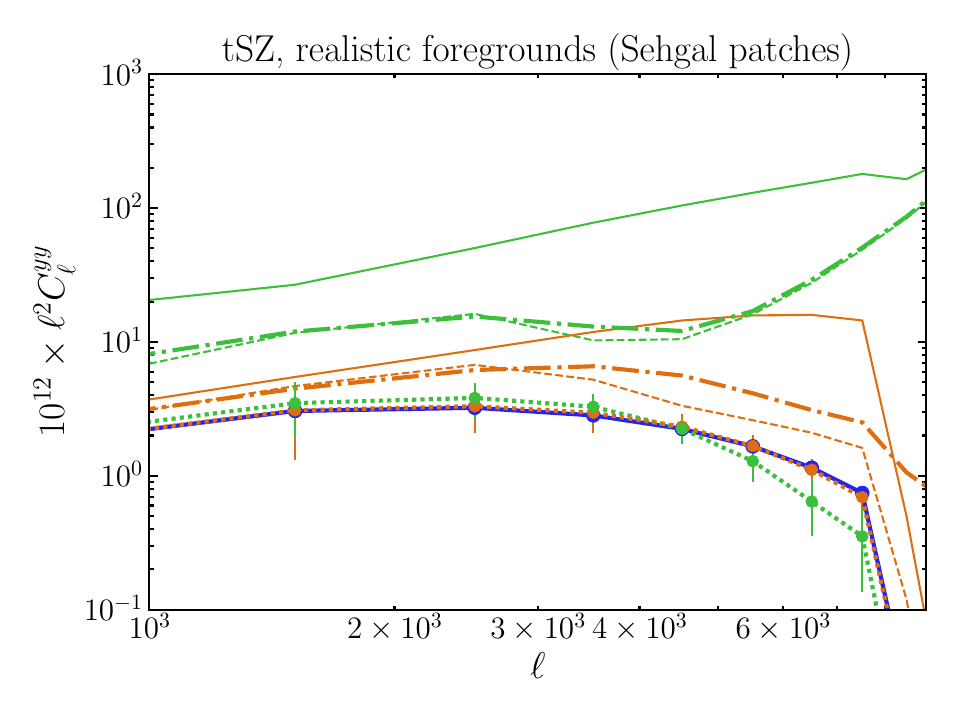}
\includegraphics[width=0.6\textwidth]{leg_autoCLsILC_withNN.pdf}
\includegraphics[width=0.2\textwidth]{leg_Xcrosstrue_NN.pdf}

\caption{The effect of including realistic correlations between the foreground and the true tSZ signal in the test sets, which have been trained on data with no such correlations. While it is no longer true that the residual is uncorrelated with the signal, the overal performance of the NN component separation as compared to ILC is qualitatively unaffected (cf. Fig.~\ref{fig:MSE_tSZ} and  the bottom right of Figs.~\ref{fig:ksz_websky} and~\ref{fig:ksz_sehgal}).  }\label{fig:restoring_fgs}
\end{figure}

We can include these correlations by always using cut-out patches of the Websky simulations that correspond to the same sky area for the foregrounds and signal. However, we only do this for the validation sets and not for the training sets that we use to train the networks. This avoids the network learning the model for the distribution of the signal through its residual correlations with the foregrounds, which it sees. Results are shown in Figure~\ref{fig:restoring_fgs} for the tSZ effect; the relative performance of the NNs as compared to the ILC is similar to the case when the foregrounds are uncorrelated with the signal. 

 \section{Demonstration: large-scale galactic B-mode polarization}\label{sec:large_scale_bmode_results}

In this section we consider the performance of our networks on large-scale B-mode CMB polarization simulations with Galactic foregrounds.  We describe our simulation sets in Section~\ref{sec:largescale_sims} and the performance of our method in Section~\ref{subsec:bmode_performance}.

\subsection{Description of simulation set}\label{sec:largescale_sims}

We consider noiseless simulated observations of the B-mode polarization of the {Galactic} mm sky, at 93, 143, 220, and 280 GHz from PySM3~\cite{2017MNRAS.469.2821T}. We include foregrounds from Galactic dust; synchrotron; and {polarized} anomalous microwave emission (AME) according to the PySM3 models [\texttt{d1},\texttt{s1},\texttt{a2}] respectively. We refer the reader to the PySM3 documentation for descriptions of these models, although we note that the dust SED is described by a modified blackbody with spatially varying temperature and spectral index and that the synchrotron SED is described by a power-law with spatially varying spectral index. The non-Gaussian templates at reference frequencies are based on the Commander dust map from the \textit{Planck} data~\cite{2016A&A...594A..10P} and the synchrotron map based on the \textit{WMAP} 27 GHz data~\cite{2013ApJS..208...20B}.

{For the CMB, in each patch we draw a random Gaussian realization of a CMB B-mode map, with each patch having a $\Lambda$CDM power spectrum calculated with \texttt{camb}~\cite{2011ascl.soft02026Lc}\footnote{\url{https://camb.readthedocs.io/en/latest/}} using cosmological parameters consistent with the \textit{Planck} 2018 cosmology~\cite{2020A&A...641A...6P}, including non-zero values of the tensor-to-scalar ratio $r<0.032$. We redraw these random patches between every epoch of training. The values of $r$ are chosen from a distribution uniform in $\log_{10}r$ with $-3<\log_{10}r<-1.5$.} 

We project the sky into 452 square patches, of side 10 degrees. We divide the sky into four quadrants in Galactic coordinates, using the Galactic center to divide the upper two quadrants from the lower two quadrants.  We then define the patches in three of the quadrants to be the training set, and the patches in the remaining quadrant to be the validation set; we augment the training dataset by including rotations around 90, 180, and 270 degrees for each patch. {We remove some patches, in particular those on the poles and those along the equator.}

\begin{figure}[t]
\includegraphics[width=0.24\textwidth]{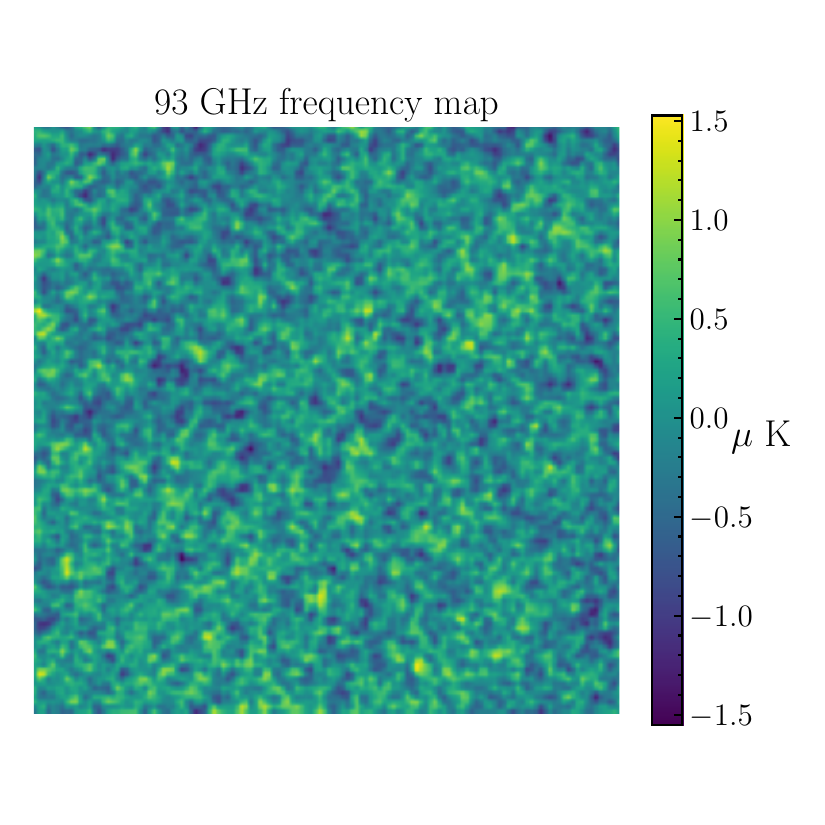}
\includegraphics[width=0.24\textwidth]{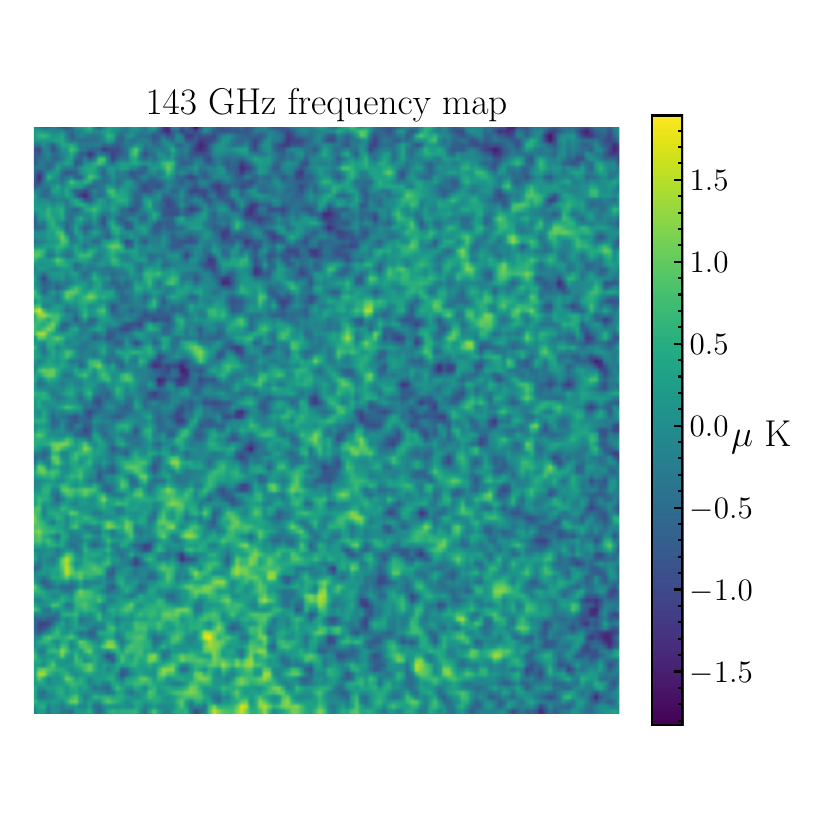}
\includegraphics[width=0.24\textwidth]{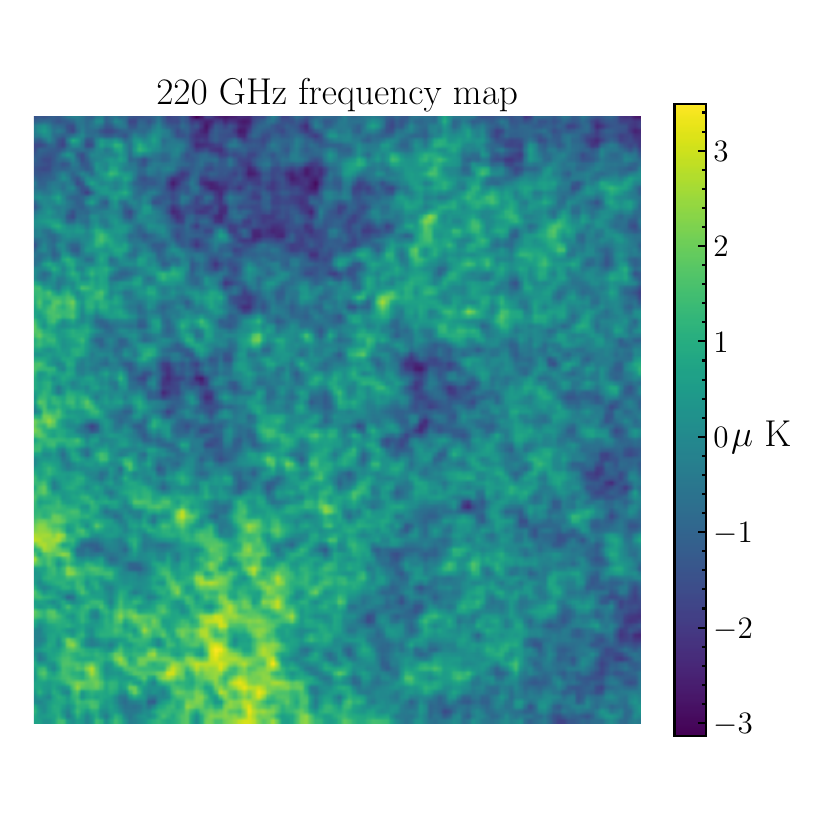}
\includegraphics[width=0.24\textwidth]{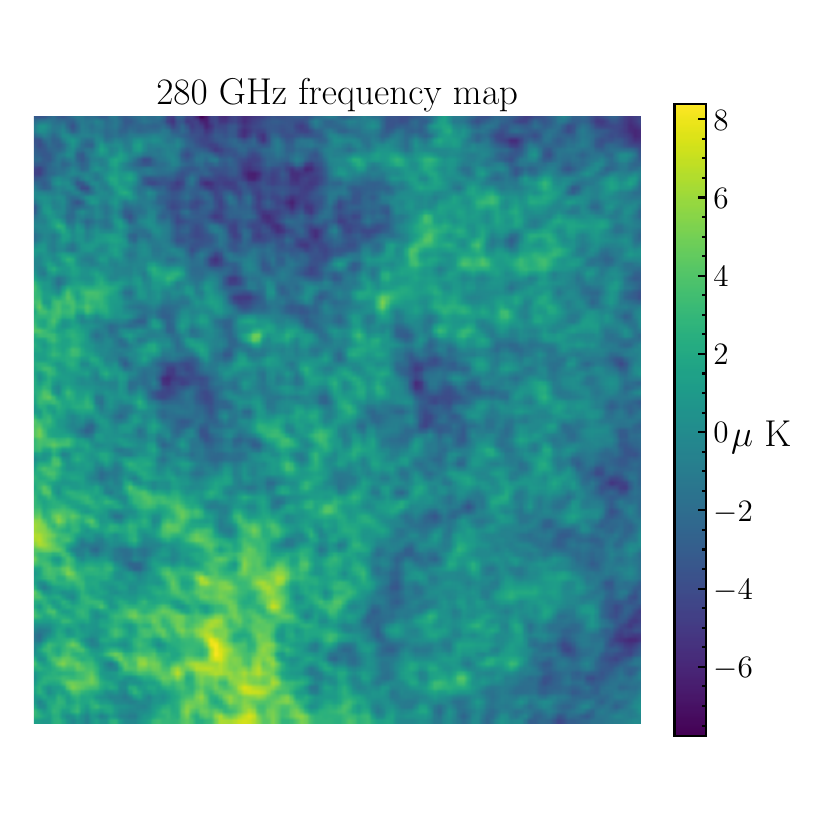}\\

\includegraphics[width=0.24\textwidth]{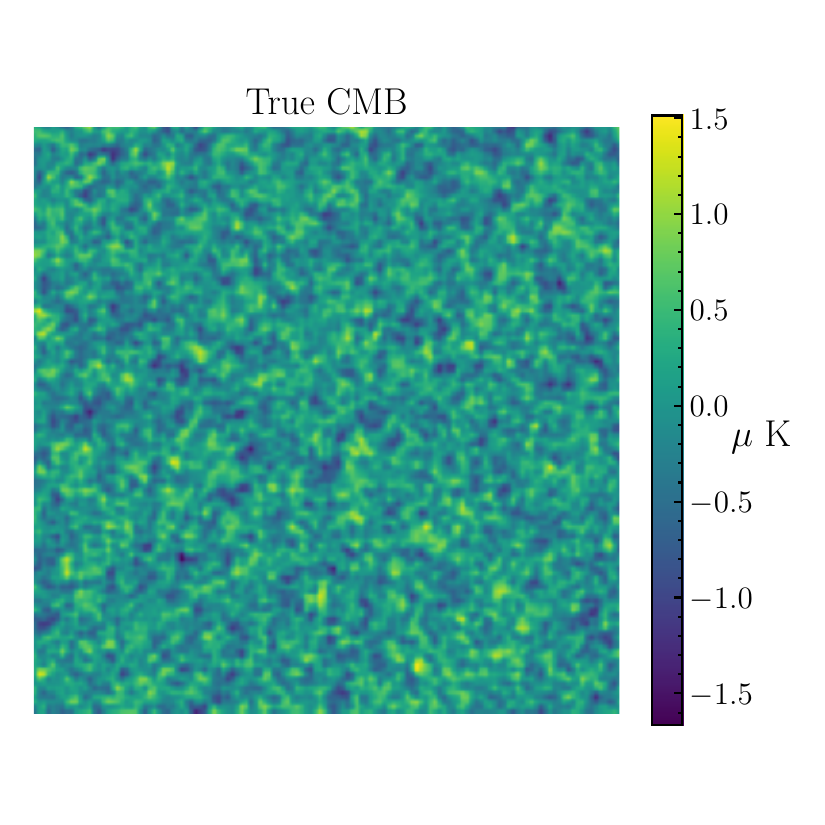}
\includegraphics[width=0.24\textwidth]{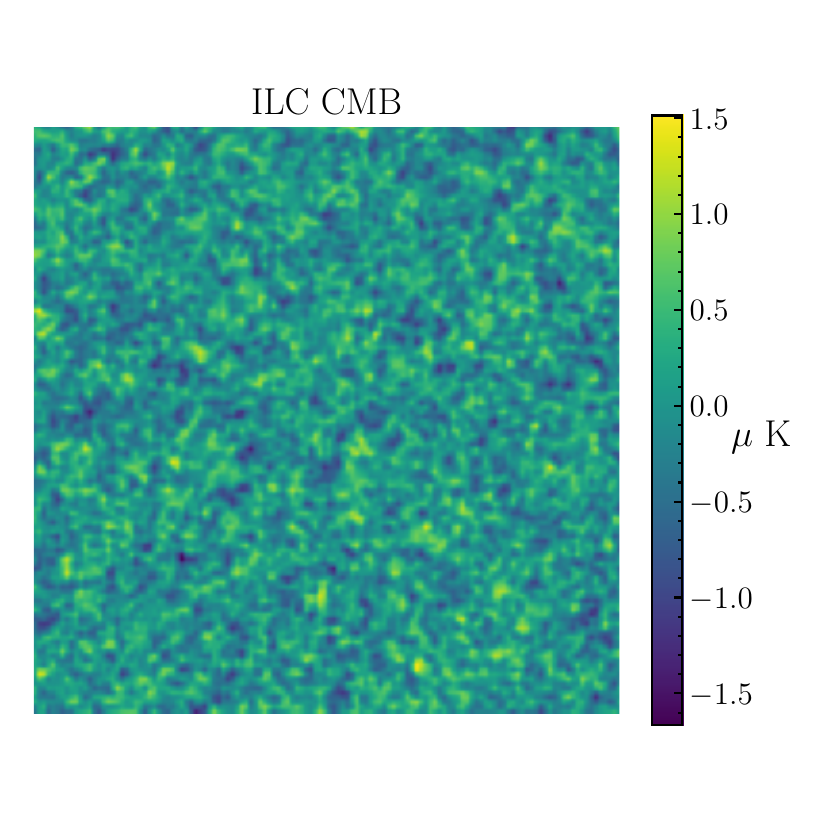}
\includegraphics[width=0.24\textwidth]{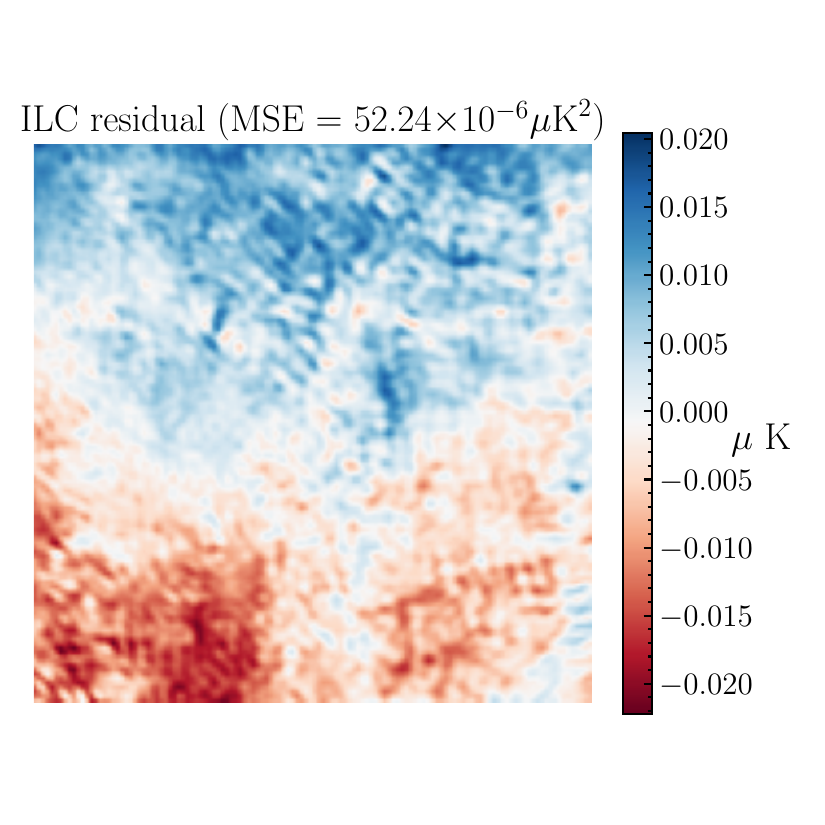}\\

\caption{A sample simulated multi-frequency large-scale polarization  measurement, {with $r=0.001$}. On the top row, we show the single-frequency maps; on the second row we show the true blackbody CMB  signal, along with the ILC determination and the residual of the ILC (defined by subtracting the truth from the ILC).   }\label{fig:sample_Bmodes}
\end{figure}

This results in 1348 training patches and 111 test patches. A sample patch is shown in Figure~\ref{fig:sample_Bmodes}. We show the single-frequency maps, along with the truth and the ILC blackbody estimate (with the covariance computed on the domain of the entire patch) and its residual with respect to the truth. In this case, we see significant anisotropy in the ILC residual, due to the anisotropy of the foreground dust component and its spectral index and temperature. The ILC is ill-suited to identifying these anisotropic features, and this is where we hope to improve in this case using our NN-based method.

\subsection{Performance}
\label{subsec:bmode_performance}

We train our networks in batches of 32 for 256 epochs with the loss function
\be
L=\sum_{i}\frac{1}{N_{i}}\sum_{p}\left(\hat {\Delta T^{\rm{ILC}}}(p)-\Delta T^{\rm{ILC}}(p)\right)^2
\ee
where the normalization factor $N_i\equiv \sum_{p} \Delta T_i^{\rm{ILC}}(p)$ penalizes the contribution of patches (labelled by $i$) that have a very large error on average (this is helpful due to the anisotropy in the signal and resulting non-uniformity of the patches).
 
\begin{figure}[h!]
 \includegraphics[width=0.49\textwidth]{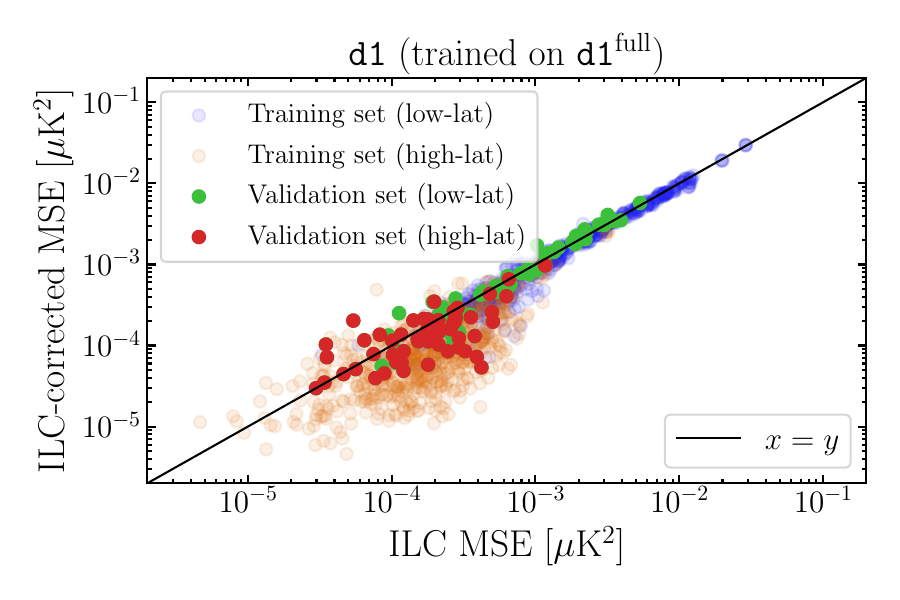}
 \includegraphics[width=0.49\textwidth]{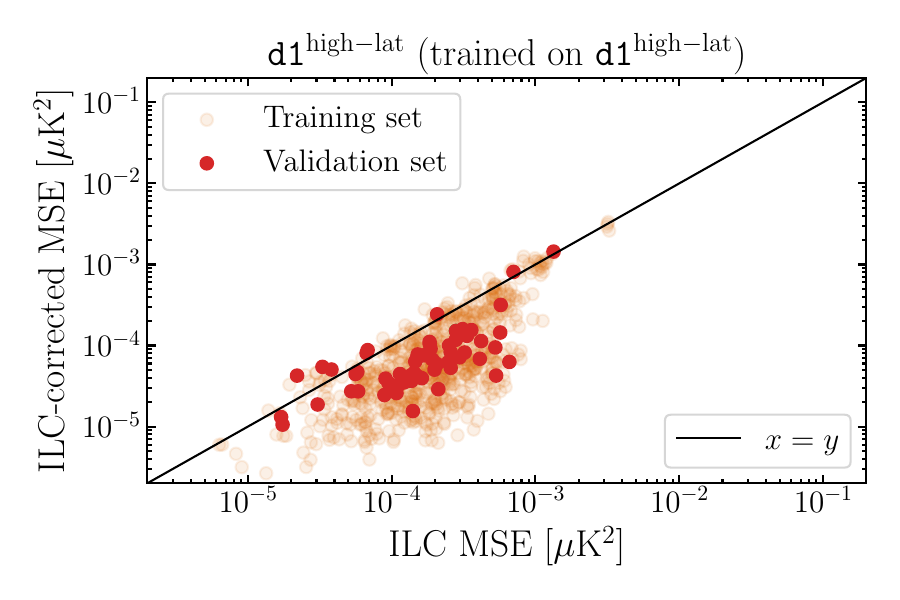}
 \caption{The per-patch MSE of the ILC versus the per-patch MSE of the ILC-corrected estimates. We indicate the line $x=y$; points under this line are indicative of a lower MSE in the ILC-corrected estimate than the ILC estimate. In all cases we define the MSE with respect to the true underlying CMB signal. On the left, we show the case where we trained the NNs on \textit{all} patches, but we indicate by the color of the points whether it is a high-Galactic-latitude ($>25^{\circ}$) or a low-Galactic-latitude patch ($<25^{\circ}$). On the right, we show the case where we trained only on high-Galactic-latitude ($>25^{\circ}$) patches.}\label{fig:scatter_dust_d1}
 \end{figure}

After training, we demonstrate our network on the \textit{test} set by comparing the per-patch MSE of our ILC-corrected residuals with the per-patch MSE of the ILC, where we define the MSE with respect to the true underlying signal. We show a scatter plot in Figure~\ref{fig:scatter_dust_d1}. We split our patches by their Galactic latitude, showing the high-Galactic-latitude and the low-Galactic-latitude patches with different colors (with the boundary being at latitude of 25$^{\circ}$). The low-Galactic-latitude patches (near the Galactic center) are much harder to clean and tend to have higher ILC MSEs. Additionally, they are harder to improve with the NN. As we are unlikely to ever clean these patches to the point where we can extract competitive cosmological information, we additionally train a network with these patches removed to see if the performance (in particular the generalization to other dust models) in the regime of interest (high Galactic latitude) can be improved when they are not included in the training set; this case is shown on the right-hand side of Figure~\ref{fig:scatter_dust_d1}.

\begin{figure}
 \includegraphics[width=0.32\textwidth]{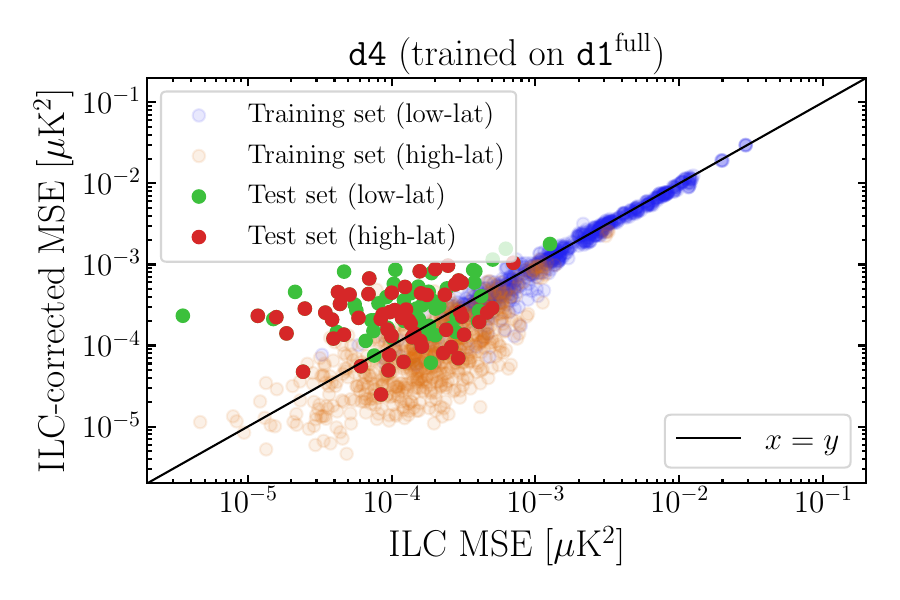}
 \includegraphics[width=0.32\textwidth]{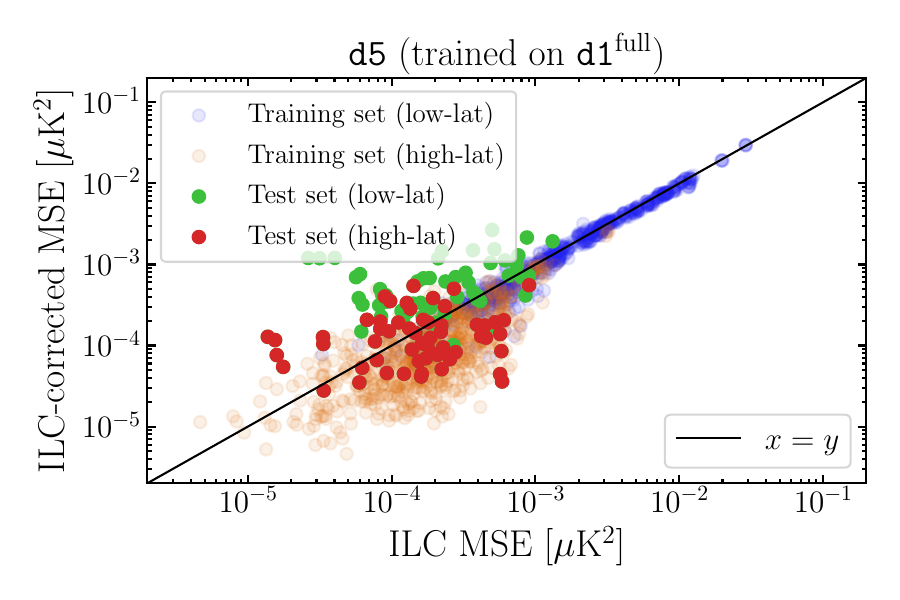}
 \includegraphics[width=0.32\textwidth]{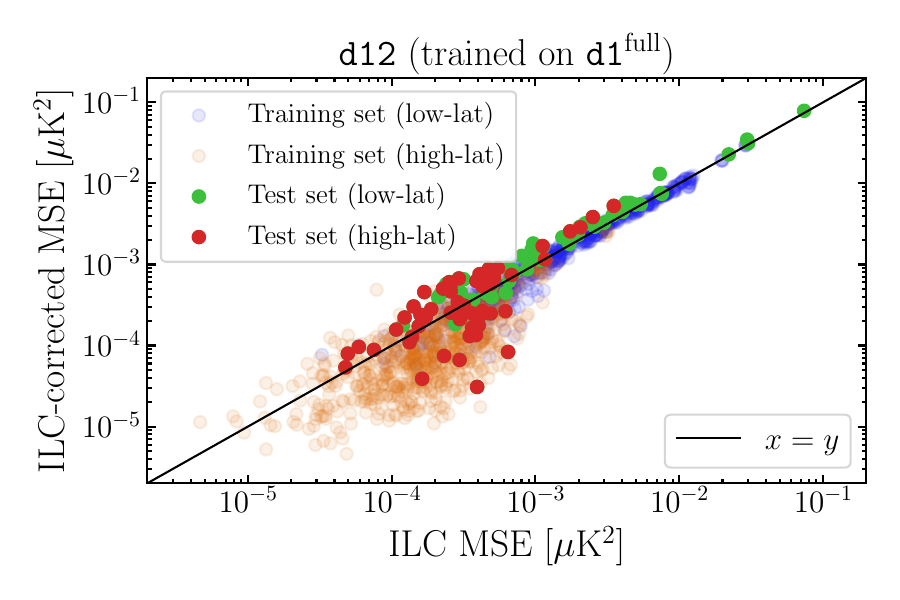}

 \includegraphics[width=0.32\textwidth]{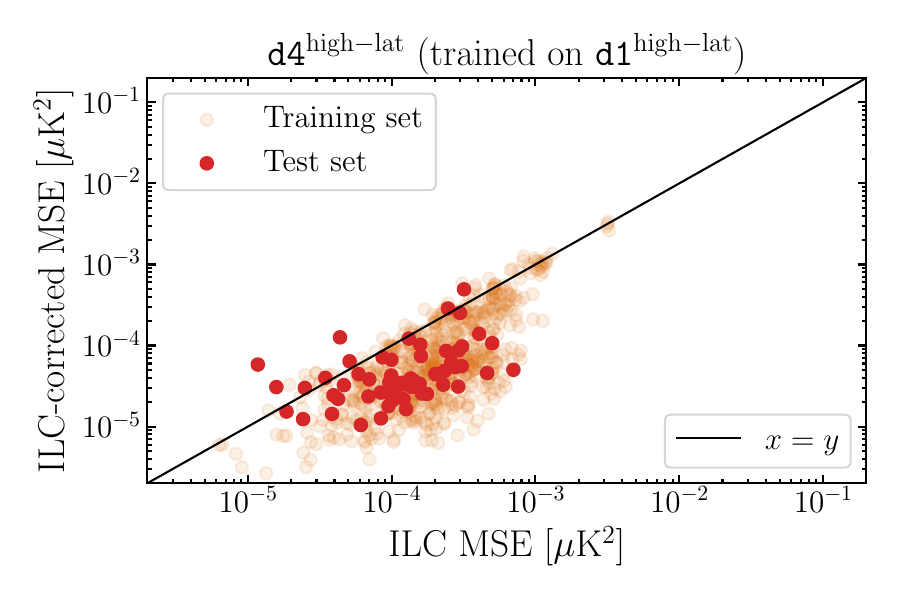}
 \includegraphics[width=0.32\textwidth]{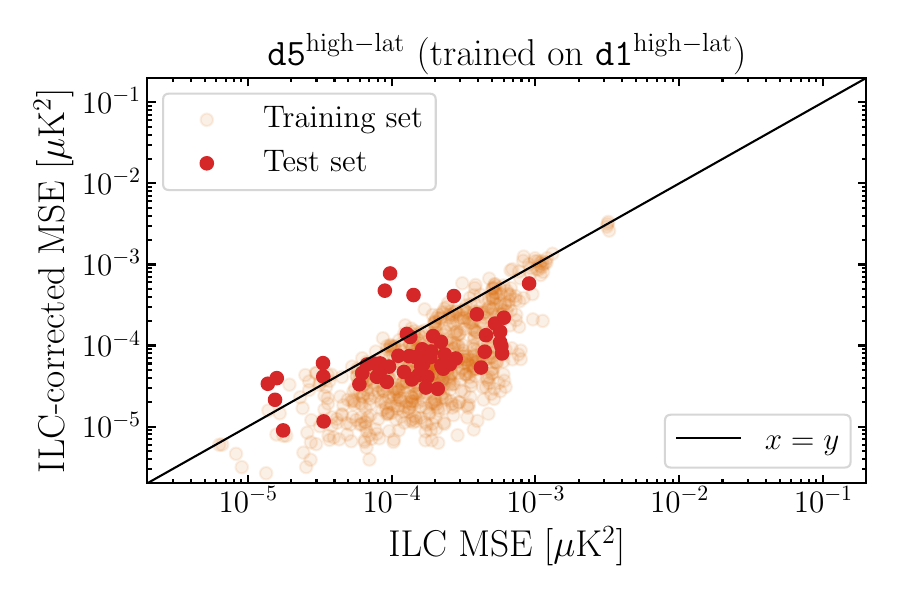}
 \includegraphics[width=0.32\textwidth]{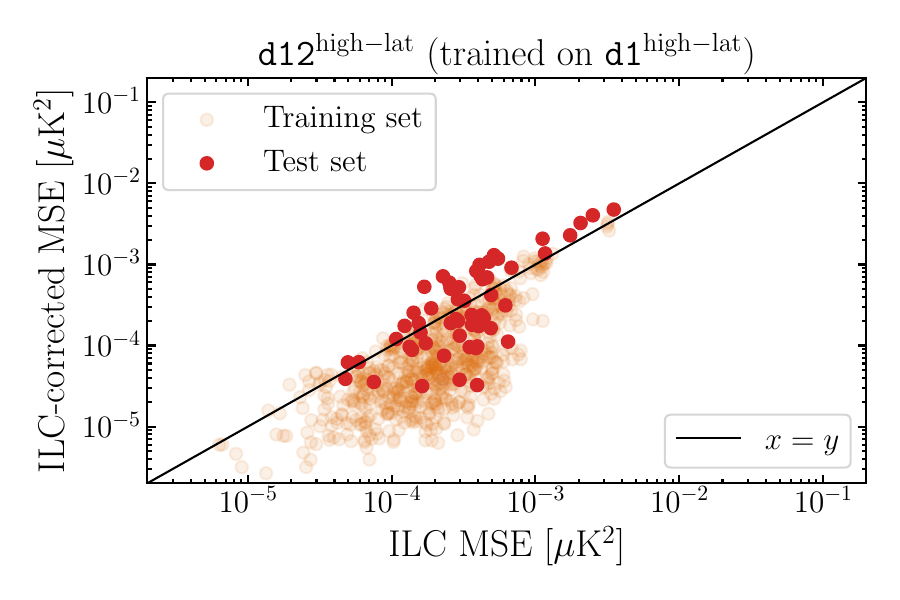}

 \caption{The per-patch MSE of the ILC versus the per-patch MSE of the ILC-corrected estimates. We indicate the line $x=y$; points under this line are indicative of a lower MSE in the ILC-corrected estimate than the ILC estimate. In all cases we define the MSE with respect to the true underlying CMB signal.  We see that the \texttt{d1}-trained model performs well even on different dust models, apart from the \texttt{d12} model in the bottom right panel. In the top panel, we have trained on a dataset including patches from the full sky; on the bottom panel, we have trained on a dataset with patches within 25$^\circ$ of the Galactic center removed. There is some improvement in the performance and generalization of the model on high-Galactic-latitude patches when we remove these patches from the training. }\label{fig:scatter_dust_dx}
 \end{figure}

We also demonstrate our method on foreground models that it has not been trained on, explicitly by creating similar datasets to the one we trained on but by replacing the \texttt{d1} model with the \texttt{d4}, \texttt{d5}, and \texttt{d12} models for the polarized dust emission in PySM3.  We continue to restrict to the quadrant that we did not train the model on. We show the resulting distributions of the per-patch MSE in Figure~\ref{fig:scatter_dust_dx}. While for the \texttt{d4} and \texttt{d5} models we continue to do better than the ILC in $>50\%$ of cases, for \texttt{d12} this number is just under $50\%$; the performance (by this metric) is summarized in Table~\ref{tab:better_than_ILC}. However, it is notable that the distributions of the ILC MSE for these models do not overlap with the distributions of the MSE of our \texttt{d1} training set; we make this clear by underplotting the corresponding points for the training set. Indeed, on the training set we see that we have preferentially trained on the large-ILC MSE loss cases. In the regime $10^{-4}<\mathrm{MSE}_{ILC}<10^{-3}$, in which our model performs best on the training set, our model in fact performs relatively well on the \texttt{d12} set. In general, the distributions on the \texttt{dx} test sets tend to follow the distributions on the \texttt{d1} test set, and the \texttt{d12} set is the only set that has significant support outside of the testing regime, possibly explaining the poorer performance here.

 \begin{table}[h!]
 \begin{tabular}{|c|c|c||c|c||c|c|}\hline
 Dataset & \multicolumn{2}{|c||}{Full set (trained on full set)}& \multicolumn{2}{|c||}{High-latitude  (trained on full set)}& \multicolumn{2}{|c|}{High-latitude  (trained on high-latitude)}\\\hline\hline
&$\mathrm{MSE} < \mathrm{MSE}^{\mathrm{ILC}}$ &$\mathrm{MSE} > \mathrm{MSE}^{\mathrm{ILC}}$ &$\mathrm{MSE} < \mathrm{MSE}^{\mathrm{ILC}}$ &$\mathrm{MSE} > \mathrm{MSE}^{\mathrm{ILC}}$ &$\mathrm{MSE} < \mathrm{MSE}^{\mathrm{ILC}}$ &$\mathrm{MSE} > \mathrm{MSE}^{\mathrm{ILC}}$ \\\hline\hline
 Validation set (\texttt{d1}) &88&23&53&4&47&10\\\hline\hline
  Test set (\texttt{d4}) &42&69&27&30&50&7\\
  Test set (\texttt{d5}) &50&61&38&19&46&11\\
  Test set (\texttt{d12}) &35&76&28&29&27&30\\\hline
 \end{tabular}
 \caption{A comparison of the performance of the NN compared to the ILC solution, as measured by number of patches with a lower per-patch MSE than that of ILC. The distributions of the $\mathrm{MSE}$ versus the $\mathrm{MSE}^{\mathrm{ILC}}$ are shown in Figure~\ref{fig:scatter_dust_dx}. The numbers indicate the number of patches with a given criterion, $\mathrm{MSE} < \mathrm{MSE}^{\mathrm{ILC}}$ indicating a patch on which the NNs perform better than ILC by this metric (i.e., have a lower MSE than the ILC), and $\mathrm{MSE} > \mathrm{MSE}^{\mathrm{ILC}}$ indicating a poorer performance than ILC.}\label{tab:better_than_ILC}
 \end{table}

\begin{figure}[h!]
 \includegraphics[width=0.32\textwidth]{pysm3_d1_nonoise_128pixels_ILCRes_patch4}
 \includegraphics[width=0.32\textwidth]{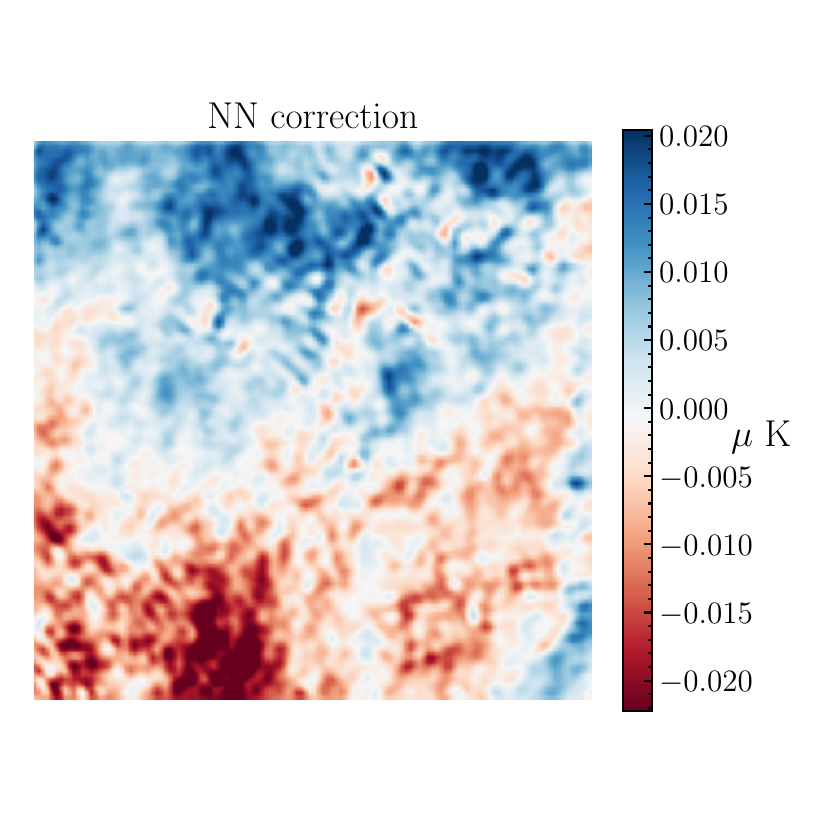}
 \includegraphics[width=0.32\textwidth]{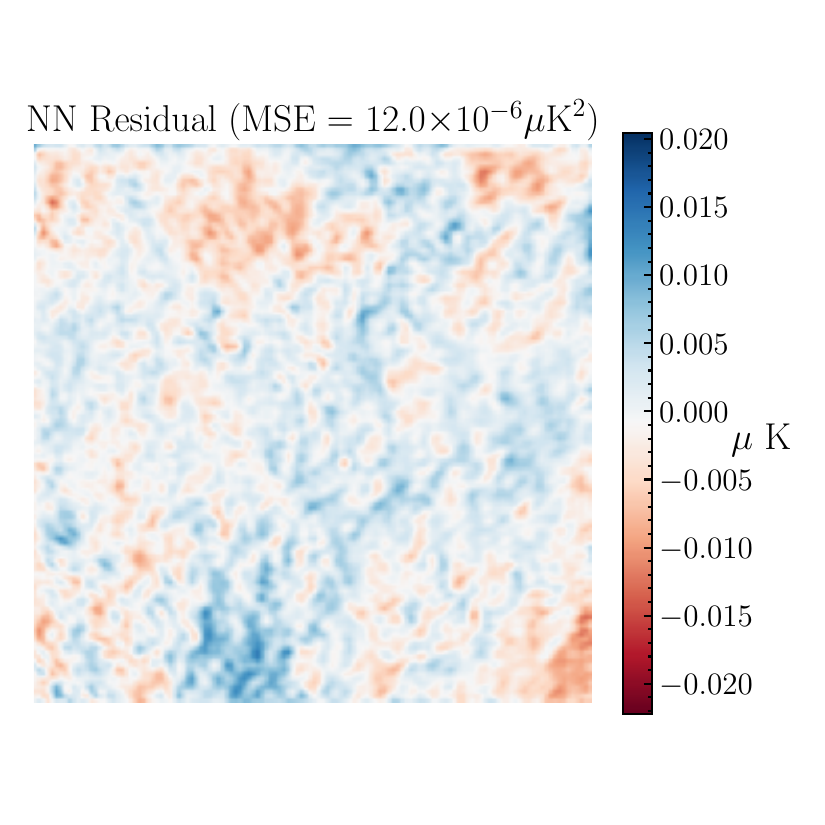}
 \caption{The performance of our NN-based method on a test \texttt{d1} patch. The ``truth'' that the NN is trying to predict is the ``ILC residual'', which is shown on the left. The prediction is the center panel; we label this ``NN correction'' as it is the correction that we subtract from the ILC solution to predict the component of interest (the CMB in this case).  By eye, it is clear that the large-scale anisotropy of the ILC residual has been successfully captured by the model. Finally, we show the residual of the NN correction on the right; this is by eye less anisotropic on large scales (also note the much smaller scale on the colorbar). This residual is also the residual on the final predicted CMB B-mode map.  Overall, we have reduced the residual from the ILC case (on the far left) by a factor of $\sim4$ (in the mean-squared-error).}\label{fig:demonstration_bmodes}
 \end{figure}

 We show explicitly the residuals for a sample patch (on which we have improved over the ILC result) from our test set in Figure~\ref{fig:demonstration_bmodes}.  It is clear that our neural network has been able to identify and correct for the anisotropic features in the foregrounds. For concreteness, we choose the same patch as we displayed in Figure~\ref{fig:sample_Bmodes}.

\section{Discussion and Conclusions}\label{sec:conclusion}

We have introduced and demonstrated a method to remove foregrounds from ILC estimates of the CMB and tSZ signals. Motivated by the reluctance to apply a complicated ML model to a signal we wish to analyze with high precision, we have proposed constructing combinations of the data that do not contain the signal of interest and only passing these to the NN.  This procedure ensures that, while the noise and foregrounds may have a complicated function applied to them, our signal is unaffected and propagates into the final map in an unbiased way. 

Our method requires training on simulations of the mm-wave sky. This is a drawback, as we must trust that in some sense our simulations are ``realistic'' and representative of the true sky. It is not clear that our method is extendable in this sense: however, we find that when we train on Galactic B-mode signals, our NNs can generalize somewhat to foreground models on which they were not trained. However, there remains motivation to train on a wider range of models and allow for significant variation in the training set, before applying such a method to the real sky.

There is further work to be done before this method is ready to be applied to real data. Various steps forward include: 

\begin{enumerate}
\item  We trained and demonstrated our model on idealized data, with the simplest assumptions for instrumental noise, and no instrumental systematics included. Further work would consider more realism, appropriate for the data one is interested in.

\item We trained on small cut-out patches of the sky, which is not the form that the data take. The data are continuous on the sphere. This leads to some options for what the best way would be to generalize our method. One option is to train on simulations that are of the same form as the data, for example by moving to larger patches, closer to the sky coverage expected from future surveys. There are various limitations to consider here: first, GPU memory prevents our models and datasets from getting very large. However, we note that it may not be strictly necessary to use a larger model when we go to larger sky areas, and that we did not use the smallest models that we could have in this demonstration. In particular, we find that when we decrease the number of layers (downsampling operations) in our UNet, we do not sacrifice performance (at least for the extragalactic case); presumably this is because the model is performing a surprisingly local operation in this case. This may not be the case for the Galactic foreground situation, where the spectral index varies slowly across a large region of the sky. However, we are hopeful that our networks may be scaled up to larger regimes; we leave a quantitative investigation to future work. 

A slightly more concerning limitation towards going to larger sky areas is the lack of multiple independent realizations of the foregrounds on the whole sky. This is a serious limitation for our current work. However we are aware that various other options exist.  Having demonstrated our network on the PySM3 simulations, a pressing follow-up work should investigate other models, perhaps with various levels of non-Gaussianity. This, too, we leave to future work. 

\item Finally, while we tried to create a ``more interpretable'' ML method to isolate signals in the sky, we are aware that it is still not clear to us what the ML model is learning. Is it learning too much the statistics of the test/training data, and minimizing them? Or is it truly learning a generic method to reduce foregrounds? Future work will look at expansions of this method to wider ranges of models, with the aim of understanding what is happening inside the black box.
\end{enumerate}

When applied to simulations on which the networks were trained, we found significant performance improvements compared to ILC, as measured by a lower MSE (and angular power spectrum, for the extragalactic components). In a low-noise case, our methods also generalize well to other simulations on which they were not trained, although unsurprisingly the inclusion of Gaussian noise  can change these improvements. While  further training in more realistic setups will be required for application to experimental data, our results represent a promising first proof-of-concept.

\begin{acknowledgements}

We thank the Scientific Computing Core staff at the Flatiron Institute for computational support. 

The Flatiron Institute is a division of the Simons Foundation. FMcC acknowledges support from the European Research Council (ERC) under the European Union’s Horizon 2020 research and innovation programme (Grant agreement No.~851274).  JCH acknowledges support from NSF grant AST-2108536, DOE grant DE-SC0011941, NASA grants 21-ATP21-0129 and 22-ADAP22-0145, the Sloan Foundation, and the Simons Foundation.

This work relied heavily on the use of numpy~\cite{harris2020array}, matplotlib~\cite{Hunter:2007}, PyTorch~\cite{2019arXiv191201703P}, healpy~\cite{Zonca2019}, and HEALPix~\cite{2005ApJ...622..759G}.

This is not an official Simons Observatory Collaboration paper.

\end{acknowledgements}

\appendix

\section{Small-scale extragalactic temperature simulations}\label{app:simulations_extragal}

We use simulations of the small-scale microwave sky from Websky~\cite{2020JCAP...10..012S} and from Sehgal et al.~\cite{2010ApJ...709..920S}. We simulate extragalactic foregrounds at frequencies corresponding to a subset of those planned for the SO Large Aperture Telescope~\cite{2019JCAP...02..056A}: 93, 145, 220, and 280 GHz.  We include the following components:

\begin{enumerate}
\item the lensed CMB; 
\item the kinetic Sunyaev--Zel'dovich (kSZ) signal; 
\item the thermal Sunyaev--Zel'dovich (tSZ) signal; 
\item the cosmic infrared background (CIB); 
\item and radio point sources~\cite{2022JCAP...08..029L}.
\end{enumerate}

For each frequency, the  CMB, kSZ, and tSZ are simply templates multiplied by the appropriate SED; for the radio point sources and CIB we use the appropriate simulated map (although, for the radio point sources and the CIB, we use the 217 GHz emission as a proxy for 220 GHz, and the 143 GHz patches as a proxy for 145 GHz).

Before adding the sources together, we project the components of each sky simulation separately into equal-area square patches with $\Npix\times\Npix$, with $\Npix=256$ pixels, with one pixel having a resolution of $0.9$ arcminutes; thus our patches are $\sim 14.7$ square degrees. We then simulate the pixel-wise temperature (in units of the CMB temperature) in a sky patch (labeled $i$) at frequency $\nu$ as follows:
\be
T^i_\nu(p) = T^i_{\mathrm{CMB}}( p) + T^i_{\mathrm{kSZ}} ( p)+ g_\nu y^i( p) + f_\nu ^i \frac{dT}{dB_\nu} I_\mathrm{CIB}^i{}_\nu(p) + \frac{dT}{dB_\nu}  I^i_{\mathrm{radio}}{}_\nu(p);
\ee
quantities that vary pixel-wise are indicated with $(p)$ while quantities without $(p)$ are constant over an entire patch. $T^i_{\mathrm{CMB}}(p) $ is the lensed CMB temperature in patch $i$; $T^i_{\mathrm{kSZ}}(p) $ is the kSZ temperature in patch $i$; $g_\nu=x\coth\lb\frac{x}{2}\rb-4$, with $x\equiv\frac{h\nu}{k_B T_{CMB}}$, is the tSZ SED; $y^i(p)$ is the Compton $y$-distortion anisotropy in patch $i$; $I_\mathrm{CIB}^i{}_\nu(p)$ is the CIB intensity in patch $i$; and $I^i_{\mathrm{radio}}{}_{\nu}(p)$ is the radio point-source intensity map in patch $i$.  As the CIB and radio intensities of Websky are provided in $\mathrm{MJy \, sr}^{-1}$, we convert them to  blackbody CMB temperature units.

\begin{figure}[t]
\includegraphics[width=0.49\textwidth]{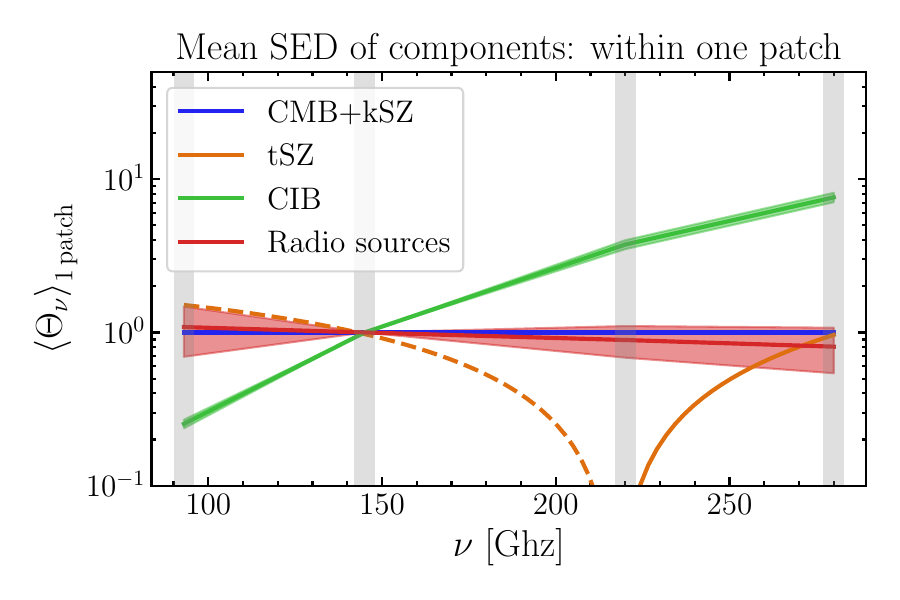}
\includegraphics[width=0.49\textwidth]{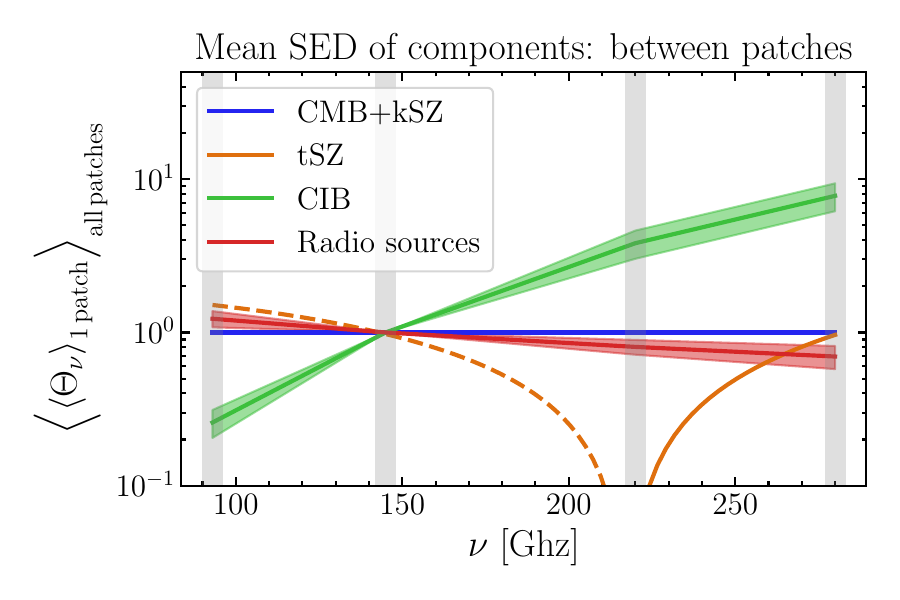}
\caption{The frequency scalings of the components. On the \textit{left}, we choose a random patch and plot the mean frequency scalings of the components, normalized at 143 GHz, i.e., the mean $\left <T(p)^\nu/T(p)^{143 \mathrm{GHz}}\right> $ where $\left<\cdot\right>$ indicates the mean over all of the pixels in the patch. We indicate the standard deviation of this quantity as a band around the mean. A non-zero standard deviation here indicates a spatially varying SED, or frequency decorrelation, within the patch. On the \textit{right}, we indicate the mean of this quantity over all of the patches. The bands on the right are the standard deviation across patches; a non-zero standard deviation here indicates inter-patch variation in the spectral indices, with intra-patch variations smeared out by the first mean operation. Note that, without our multiplication by the random CIB scaling factor $f_\nu$, the CIB standard deviation would be much closer to zero.  Also note that there is no standard deviation on the tSZ and CMB+kSZ lines, as these have constant SEDs.}\label{fig:CIB_SED}
\end{figure}

In our simulations, we include a random factor $ f_\nu ^i$ which multiplies the CIB patches. At each $i$ and $\nu$, we draw $f_\nu^i$ from a uniform distribution between $\left[0.75,1.25\right]$. This allows us to include a small level of model uncertainty in the CIB simulations, as this amounts to varying the frequency scaling (or SED) of the CIB. Thus we train on simulations with a range of CIB SEDs, to prevent our model assuming that the SED is the (near-)constant SED of the Websky simulation maps, as in reality we do not know with such high certainty the true SED of the CIB at these frequencies. The distribution of the resulting CIB SED is shown in Figure~\ref{fig:CIB_SED}.

For the cases in which we include noise, we add Gaussian white noise at levels corresponding to those expected from SO; we use the ``goal'' configuration of  Table~1 of~\cite{2019JCAP...02..056A}: the noise levels are $\{5.8,6.3,15,37\}\, \mathrm{\mu K-arcmin}$ respectively, with corresponding beam full width at half maxima of $\{2.2,1.4,1.0,0.9\}\, \mathrm{arcmin}$. We convolve all of our maps to a common Gaussian beam of 0.9 arcminutes; thus for the final maps, the appropriate white noise has a power spectrum
\be
N_\ell^\nu = N^\nu_{\mathrm{white}}\frac{B_\ell(\theta_{0.9\,\mathrm{arcmin}})}{B_\ell(\theta_{\mathrm{FWHM}})}
\ee
where $N^\nu_{\mathrm{white}}$ is the noise in $\mathrm{\mu K}^2$ (this can be calculated easily from the noise $n^\nu$ in $\mathrm{\mu K-arcmin}$ by $N^\nu_{\mathrm{white}} = \left(\frac{\pi}{180\times 60}\times n^\nu\right)^2$), and the Gaussian beam convolving functions $B_\ell(\theta)$ are given by 
\be
B_\ell(\theta)=\exp\lb\frac{-\lb\ell\lb\ell+1\rb\rb}{2}\frac{\theta^2}{8\ln 2}\rb,
\ee
where $\theta$ is the angle in radians.

The final beam-convolved sky temperature is
\be
T^i_\nu(p){}^{\mathrm{beam-convolved}} = B_\ell(\theta_{0.9 \, \mathrm{arcmin} }) T^i_\nu(p) +N^i_\nu(p),
\ee
where $N^i_\nu(p)$ is a random map with power spectrum given by $N_\ell^\nu$, and is not included in the noiseless simulations.

\section{Neural network architecture and training}\label{app:architecture}

We consider a UNET~\cite{2015arXiv150504597R}-like architecture; we refer the reader explicitly to Ref.~\cite{2015arXiv150504597R} for a description of UNET. Our network takes in input in   $\Npix\times \Npix$ input in $N_{\mathrm{freq}}$ channels and outputs $\Npix\times \Npix$ input in 1 channel. 

In general, UNET downsamples the input with an encoder, which consecutively downsamples the input, increasing the number of channels by a factor of 2 at each downsampling operation and decreasing the number of pixels by a factor of 2. We modify this by allowing our network to take as a hyperparameter a number $N_{\mathrm{channels}}^{(1)}$, which describes the number of channels at the end of the first downsampling operation; after this, we always increase the number of channels by a factor of 2 at each downsampling operation. We generally use $N_{\mathrm{channels}}^{(1)}=32$. We also use a hyperparameter $N_{\mathrm{downsamples}}$ which describes the depth of the downsampling operation; in general, there is an $\Npix$-dependent maximum value which $N_{\mathrm{downsamples}}$ can take, which corresponds to the case where the final downsampling operation decreases the number of pixels to 1. Using this maximum value of $N_{\mathrm{downsamples}}$ allows the UNET to use information about correlations on the entire scale of the patch; in practice, however, we find it is not always necessary to use this maximum value (and using lower values allows us to have a smaller model, as the size of the model increases exponentially in $N_{\mathrm{downsamples}}$ due to the number of channels doubling at each downsampling operation). We downsample $\Npix$ by using a $2\times 2$ Average(mean)-Pool operation (as in the standard UNET).

Between each downsampling operation, we apply consecutively  ${N_{\mathrm{conv}}}$ convolution operations, with each convolution followed by an activation function. As in the standard UNET, we always take ${N_{\mathrm{conv}}}=2$ and allow the size of the convolution kernels to be 3. Our activation function is a rectified linear unit (ReLU) .

After the downsampling is finished, the UNET upsamples the results with a decoder. We follow the standard upsampling procedure of UNET, where the number of pixels is doubled and the number of feature channels halved at each upsampling operation using a $3\times 3$ convolution; after each upsampling step the result is concatenated with the output of the encoder with the corresponding number of pixels. After each upsampling operation, we again apply consecutively ${N_{\mathrm{conv}}}$ $3\times3$ convolution operations, with each convolution followed by an activation function. Where necessary, we pad the pixels before concatenation of the encoder and decoder output so that we are always doubling the number of pixels.

Before the final upsampling to reach the input number of pixels $\Npix\times\Npix$, we enforce that the output has only one channel. We finally apply a $3\times3$ convolution to the final output. 

After every convolutional operation we also apply a batch normalization operation. We work with a batch size of 32.  The network used in the case of the B-mode component separation is described in the Table~\ref{tab:architecture}. For the temperature component separation, we use a similar architecture, except the input has $(4,256,256)$ dimensions,  and encoder reduces $A$ times until $X_A$ has dimension $A,B,C$.

We train our networks with the Adam optimizer~\cite{2014arXiv1412.6980K}, with an initial learning rate of $10^{-3}$. We implement all our networks in \texttt{PyTorch}~\cite{NEURIPS2019_9015}.

\begin{table}[]
\begin{tabular}{|c |}\hline
Input : ($ 4, 128,128$)\\\hline
\end{tabular}
    \begin{tabular}{|c|c|c|}\hline
        \multicolumn{3}{|c|}{Encoder}\\\hline\hline
         Conv ($32\times3\times3$) &($ 32, 128,128$)&$X_1$\\\hline
         ReLu &($32, 128,128$)&\\\hline
         BatchNorm&($32, 128,128$)&\\\hline
         Conv ($32\times3\times3$) &($32, 128,128$)&\\\hline
         ReLu &($32,  128,128$)&\\\hline
         BatchNorm&($32,128,128$)&\\\hline
         AvgPool ($2\times2$)&($32,64,64$)&$X_2$\\\hline\hline
         Conv ($64\times3\times3$) &($ 64,64,64$)&\\\hline
         ReLu &($64, 64,64$)&\\\hline
         BatchNorm&($64, 64,64$)&\\\hline
         Conv ($64\times3\times3$) &($64, 64,64$)&\\\hline
         ReLu &($64, 64,64$)&\\\hline
         BatchNorm&($64, 64,64$)&\\\hline
         AvgPool ($2\times2$)&($64,32,32$)&$X_3$\\\hline\hline
         Conv ($128\times3\times3$)  &($128,32,32$)&\\
         $\cdots$&$\cdots$&$\cdots$\\
        
        AvgPool ($2\times2$)&($512,4,4$)&$X_6$\\\hline\hline
        Conv ($1024\times3\times3$)  &($1024,4,4$)&\\\hline
         ReLu &($1024,4,4$)&\\\hline
         BatchNorm&($1024,4,4$)&\\\hline
         Conv ($1024\times3\times3$) &($1024,4,4$)&\\\hline
         ReLu &($1024,4,4$)&\\\hline
         BatchNorm&($1024,4,4$)&\\\hline
         AvgPool ($2\times2$)&($2048,2,2$)&$X_7$\\\hline
    \end{tabular}
       \begin{tabular}{|c|c|}\hline
        \multicolumn{2}{|c|}{Decoder}\\\hline\hline
         Inv-Conv ($1024\times3\times3$) &($ 1024,4,4$)\\\hline
         Cat $X_6$&($  2048,4,4$)\\\hline\hline
         Conv ($1024\times3\times3$) &($  1024,4,4$)\\\hline
         ReLu &($  1024,4,4$)\\\hline
         BatchNorm&($  1024,4,4$)\\\hline
         Conv ($1024\times3\times3$) &($  1024,4,4$)\\\hline
         ReLu &($  1024,4,4$)\\\hline
         BatchNorm&($  1024,4,4$)\\\hline\hline
         Inv-Conv  ($512\times3\times3$)&($512,8,8$)\\\hline
         Cat $X_5$&($  1024,8,8$)\\\hline\hline
         $\cdots$&$\cdots$\\\hline\hline
        Inv-Conv ($32\times3\times3$) &($  32,128,128$)\\\hline
        Cat $X_1$&($  64,128,128$)\\\hline\hline
         Conv ($32\times3\times3$) &($ 32,128,128$)\\\hline
         ReLu &($  32,128,128$)\\\hline
         BatchNorm&($   32,128,128$)\\\hline
         Conv ($1024\times3\times3$) &($   32,128,128$)\\\hline
         ReLu &($   32,128,128$)\\\hline
         BatchNorm&($  32,128,128$)\\\hline
         \multicolumn{2}{c}{ }
       \vspace{2em}\\\hline
        \multicolumn{2}{|c|}{Final}\\\hline\hline
         Conv ($1\times128\times128$) &  ($1\times128\times128$)\\\hline
    \end{tabular}
    \caption{The neural network architecture for the large-scale B-mode separation, which takes as input an observation with $128\times128$ pixels and 4 frequency channels. ``Conv($x\times y\times y$)'' refers to a convolution of kernel $y\times y$ with $x$ channels. All convolutions have stride 2. ``ReLu'' refers to the rectified linear unit activation function. ``AvgPool ($x\times x$)'' refers indicates downsampling by taking the mean in each $x\times x$ window. ``Inv-Conv ($x\times y\times y$) '' refers to upsampling by taking an inverse convolution with $x$ channels and kernel size $y\times y$. ``Cat $X_A$'' refers to concatenation with the object $X_A$ indicated in the ``encoder'' table. Where necessary, we pad objects with zeros when upsampling them with InvCov so that they can be concatenated with the appropriate $X_A$ object.}
    \label{tab:architecture}
\end{table}

\bibliography{references}

\end{document}